\documentstyle[graphicx]{mn}
\def\Mfunction{{}}
\input{psfig}
\onecolumn

\def\Eq#1{{Eq.~(\ref{e:#1})}}

\def\Fig#1{{Fig.~(\ref{f:#1})}}

\def\be{\begin{equation}}
\def\ee{\end{equation}}
\def\ba{\begin{eqnarray}}
\def\ea{\end{eqnarray}}

\def\EQN#1{\label{e:#1}}
\def\Tab#1{{Table~(\ref{t:#1})}}

\def\kms{{\rm km\,s}^{-1}}

\def\spose#1{\hbox to 0pt{#1\hss}}
\def\lta{\mathrel{\spose{\lower 3pt\hbox{$\mathchar"218$}}
     \raise 2.0pt\hbox{$\mathchar"13C$}}}
\def\gta{\mathrel{\spose{\lower 3pt\hbox{$\mathchar"218$}}
     \raise 2.0pt\hbox{$\mathchar"13E$}}}

\begin{document}

\title{{\bf Stellar Dynamics in the Galactic centre: }\\
Proper Motions and Anisotropy }
\author[R. Genzel, C.Pichon, A.Eckart, O.Gerhard
 and T.Ott]{R. Genzel$^{1}$, C.Pichon$^{2,3}$, A.Eckart$^{1}$, O.E.Gerhard$^{2}$ and
T.Ott$^{1}$ \\
$^{1}$ Max-Planck Institut f\"{u}r extraterrestrische Physik, Garching,
Germany \\
$^{2}$ Astronomisches Institut, Universit\"{a}t Basel, Switzerland \\
$^{3}$ Observatoire de Strasbourg, 11 rue de l'Observatoire, Strasbourg,
France}
\date{\today}
\maketitle

\begin{abstract}
We report a new analysis of the stellar dynamics in the Galactic centre,
based on improved sky and line-of-sight velocities for more than one hundred
stars in the central few arcseconds from the black hole candidate SgrA*. The
main results are:

$\bullet$ Overall the stellar motions do not deviate strongly from isotropy.
For those 32 stars with a determination of all three velocity components the
absolute, line of sight and sky velocities are in good agreement, consistent
with a spherical star cluster. Likewise the sky-projected radial and
tangential velocities of all 104 proper motion stars in our sample are also
consistent with overall isotropy.

$\bullet$ However, the sky-projected velocity components of the young,
early type stars in our sample indicate significant deviations from
isotropy, with a strong radial dependence. Most of the bright HeI
emission line stars at separations from 1" to 10'' from SgrA* are on
tangential orbits.  This tangential anisotropy of the HeI stars and
most of the brighter members of the IRS16 complex is largely caused by
a clockwise (on the sky) and counter-rotating (line of sight, compared
to the Galaxy), coherent rotation pattern. The overall rotation of
the young star cluster probably is a remnant of the original angular
momentum pattern in the interstellar cloud from which these stars were
formed.

$\bullet$ The fainter, fast moving stars within
$\approx{}1^{\prime\prime}$ from SgrA* appear to be largely moving on
radial or very elliptical orbits. We have so far not detected
deviations from linear motion (i.e. acceleration) for any of
them. Most of the SgrA* cluster members also are on clockwise
orbits. Spectroscopy indicates that they are early type stars. We
propose that the SgrA* cluster stars are those members of the early
type cluster that happen to have small angular momentum and thus can
plunge to the immediate vicinity of SgrA*.

$\bullet$ We derive  an anisotropy-independent estimate of  the Sun-Galactic
centre   distance  between 7.8   and  8.2 kpc,  with   a  formal statistical
uncertainty of $\pm 0.9 {\ {\rm kpc}}$.

$\bullet$    We explicitly include  velocity  anisotropy   in estimating the
central mass distribution. We show how  Leonard-Merritt and Bahcall-Tremaine
mass estimates  give systematic offsets in  the inferred mass of  the central
object  when  applied to  finite concentric  rings  for  power law clusters.
Corrected Leonard-Merritt  projected  mass  estimators  and  Jeans  equation
modelling confirm previous conclusions (from isotropic models) that a compact
central mass concentration (central density $\geq$10$^{{\ 12.6 }}$M$_{{\sun}
} $ pc$^{{\ -3}}$) is present and dominates the potential between 0.01 and 1
pc.    Depending on the modelling method used the derived central mass ranges 
between  2.6 and $3.3 \times     10^{{\ 6}}M_{{\sun}   }  $  for
R$_{{\sun}}=8.0 {\ {\rm kpc}}$.
\end{abstract}

\section{Introduction}

\label{s:intro} 

High spatial resolution observations of the motions of gas and stars have in
the past decade substantially strengthened the evidence that central dark
mass concentrations reside in many and perhaps most nuclei of nearby
galaxies (Kormendy and Richstone 1995, Magorrian et al. 1998, Richstone et
al. 1998). These dark central masses are very likely massive black holes.
The most compelling evidence for this assertion comes from the dynamics of
water vapor maser cloudlets in the nucleus of NGC 4258, and from the stellar
dynamics in the centre of our own Galaxy (Greenhill et al. 1995, Myoshi et
al. 1995, Eckart and Genzel 1996, 1997, Genzel et al. 1997, Ghez et al.
1998). In both cases the gas and stellar dynamics indicate the presence of
an unresolved central mass whose density is so large that it cannot be
stable for any reasonable length of time unless it is in form of a massive
black hole (Maoz 1998).\newline
{\vskip0.21cm}The case of the Galactic centre is particularly intriguing, as
it is very close (8 kpc). With the highest spatial resolution observations
presently available in the near-infrared ($\leq $0.1''), spatial scales of a
few light days can be probed. Measurements of both line-of-sight velocities
(through Doppler shifts in spectral lines) and sky/proper motions are
available and pose very strong constraints on the central mass
concentration. The following results have emerged.

$\bullet $ The mean stellar velocities (or velocity dispersions) follow a
Kepler law ($\langle $v$^{{\ 2}}\rangle $ $\propto $ R$^{{\ -1}}$) from
projected radii R from 0.1'' to $\approx ${}20'', providing compelling
qualitative evidence for the presence of a central point mass (Sellgren et
al. 1990, Krabbe et al. 1995, Haller et al. 1996, Genzel et al. 1996, 1997,
Eckart and Genzel 1996, 1997, Ghez et al. 1998).

$\bullet$ The positions of the dynamical centre (maximum velocity
dispersion) and of the maximum stellar surface number density agree with the
position of the compact radio source SgrA* (size less than a few AU, Lo et
al. 1998, Bower and Backer 1998) to within $\pm$0.1'' (Ghez et al. 1998).

$\bullet $  Projected  mass estimators and Jeans   equation modeling of  the
stellar velocity data indicate that the  central mass ranges between 2.2 and
$ 3\times 10^{{\ 6}} M_{\odot }$. It has  a mass to  luminosity ratio of $
M/L>100 M_{\sun }/L_{\sun }$ and a density $\geq 2\times 10^{{\ 12}} M_{\sun
} {\ {\rm pc}}^{{\ -3}}$ (Genzel et al. 1997, Ghez et al. 1998).\newline

The Galactic  centre mass modeling has so far assumed that the stellar velocity
ellipsoid is isotropic.  An initial comparison  of  line-of-sight and proper
motion  velocity dispersions  indeed  suggests  that   there are  no  coarse
deviations from isotropy (Eckart and Genzel 1996, 1997).  However, to make a
more detailed assessment,  it is necessary  to obtain more accurate  stellar
motions than were available two years ago.  These improved motions -for
line-of-sight and sky  components - are  now  in  hand  and  will be
analysed in the present paper.\newline {\vskip0.21cm}

\section{\bf Observations and Data Analysis}

\begin{figure}
\centering \includegraphics[width=0.5\columnwidth,clip=true,angle=-90]{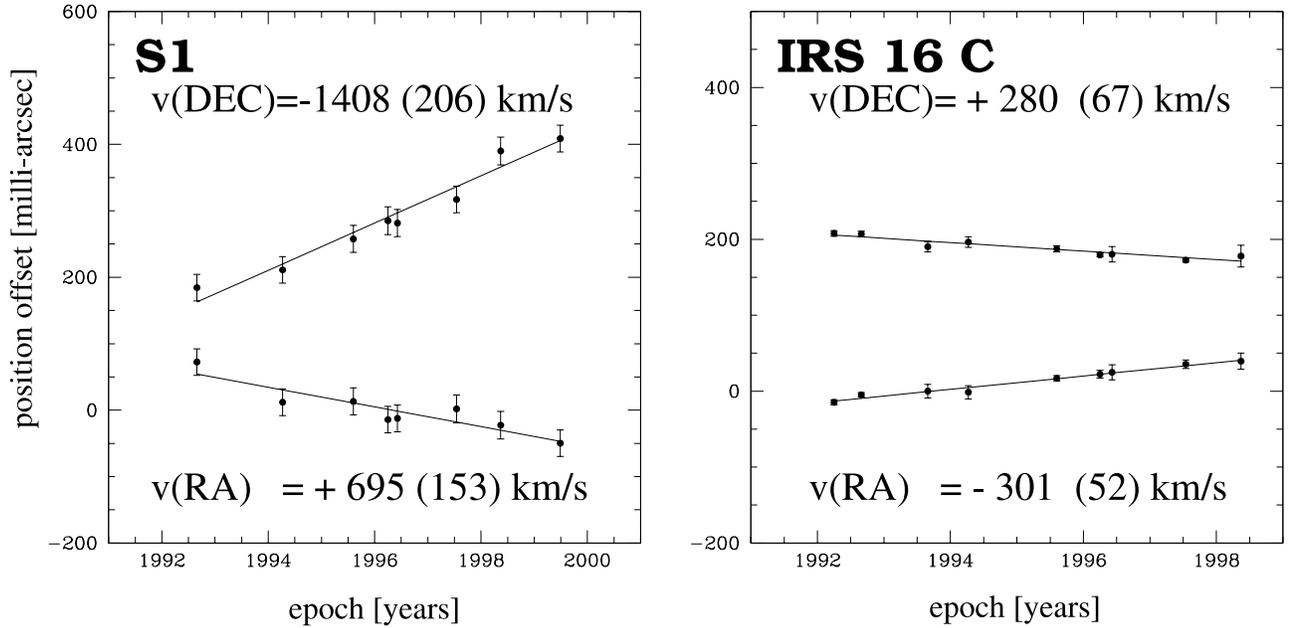}
\caption{ Examples of the proper motions derived from the 1992 to 1998 data
NTT data sets  obtained with the MPE  SHARP  camera. Shown are  position-time
plots for source S1 close  to SgrA* (left) and  the bright HeI emission line
star   IRS16 C (right).  The two  panels show x-(=RA)  and y-(=Dec) position
offsets, along  with the best-fitting  proper motions (straight  lines). For
each  epoch the  average   and 1$\protect\sigma $  dispersion   of  a number
position measurements are plotted. }
\label{f:genzel1}
\end{figure}

\begin{figure}
\centering \includegraphics[width=0.95\columnwidth,clip=true]{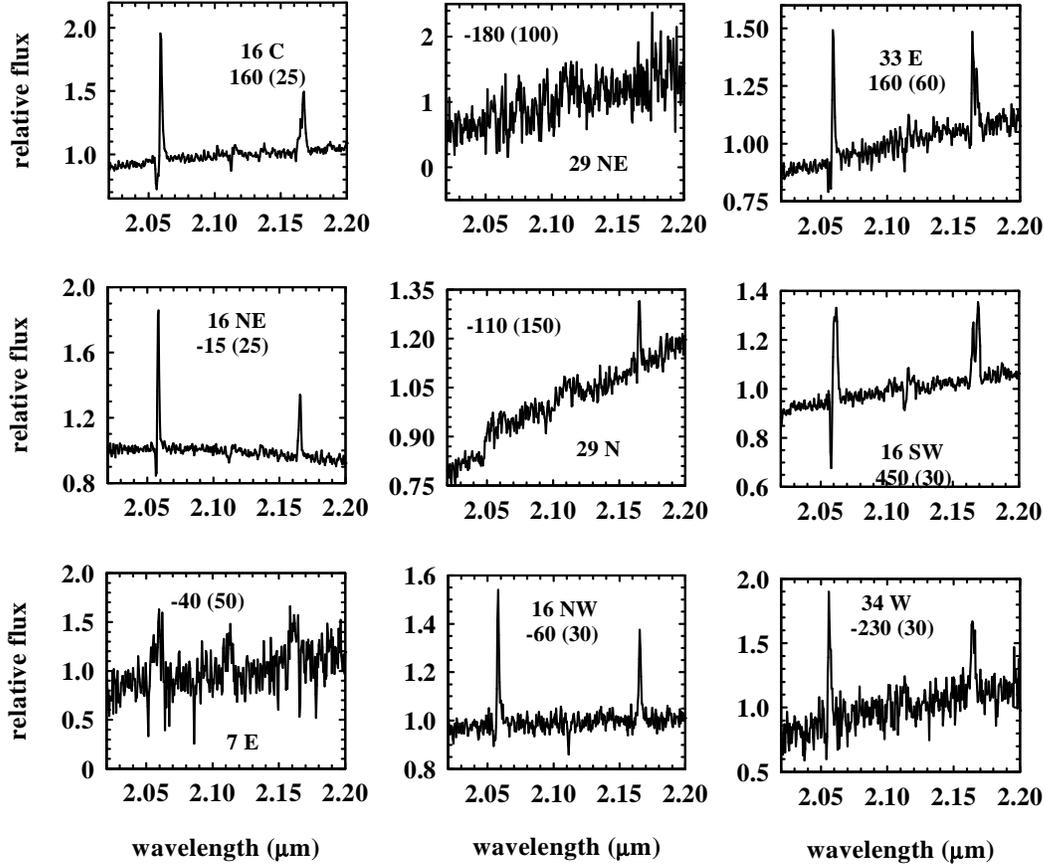}
\vspace{-7cm}
\caption{ K-band spectra of early type stars in the central $\approx
{}10^{\prime \prime }$ obtained in spring 1996 with the MPE 3D spectrometer
on the ESO-MPG 2.2m telescope . The spectral resolving power is $\protect
\lambda $/$\Delta \protect\lambda =2000$, Nyquist sampled at twice that
resolution. Most of the spectra were obtained by subtracting from the
`on-star` spectrum ($0.3^{{\prime \prime }}$ to $1.2^{\prime \prime }$
aperture) an `off-star` spectrum scaled to the same area, in order to remove
local nebular emission. The name is given for each star,
 as is the stellar velocity in km/s, with the 1 $
\protect\sigma $ uncertainty in parentheses. }
\label{f:genzel2a}
\end{figure}
\begin{figure}
\centering \includegraphics[width=0.95\columnwidth,clip=true]{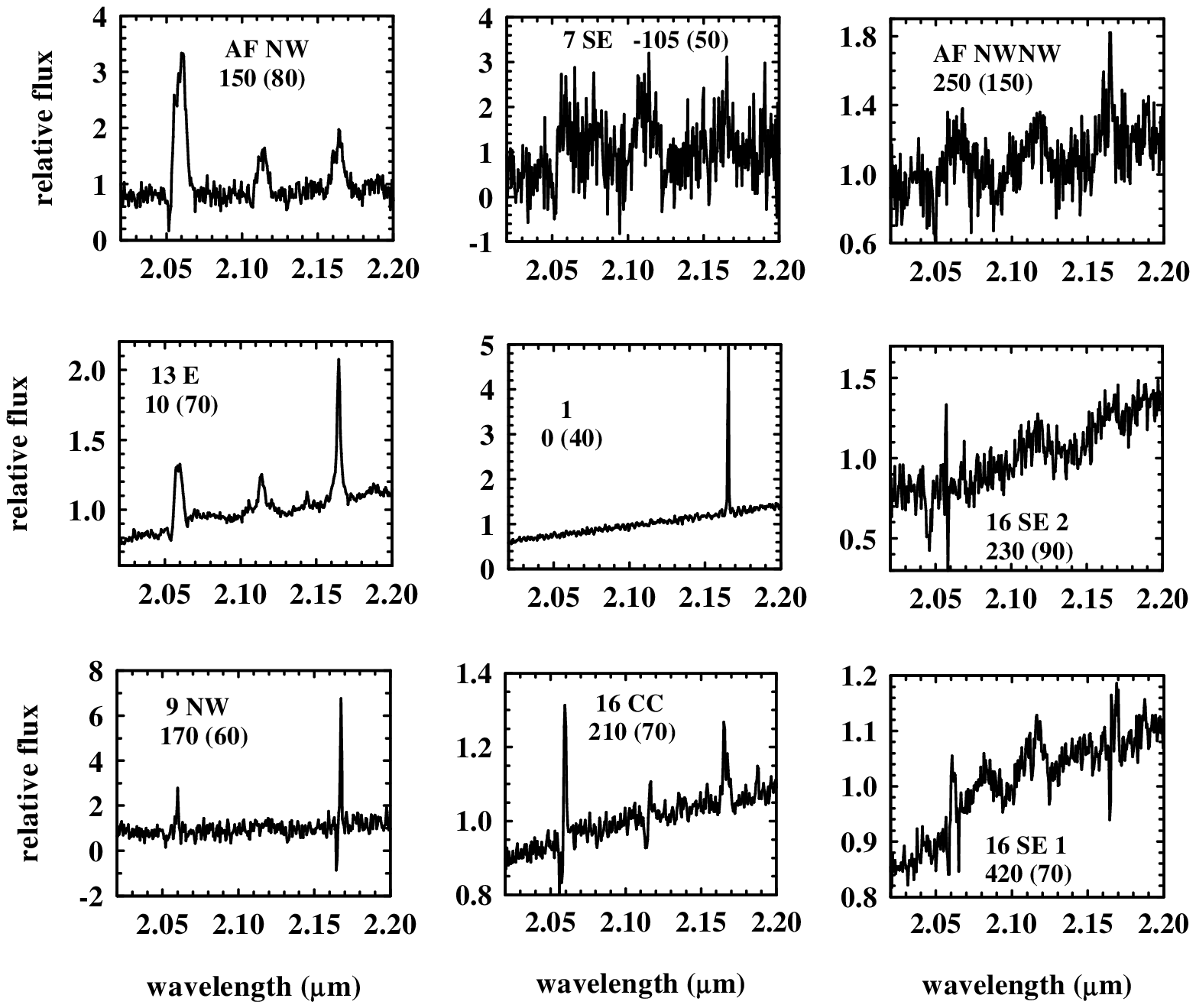}
\caption{see \Fig{genzel2a}}
\label{f:genzel2b}
\end{figure}

\subsection{\bf Proper Motions}

In our earlier papers (Eckart et al. 1992, 1993, 1994, 1995) we have
described the data acquisition and reduction that allowed us to obtain
stellar positions with a precision of $\approx ${}10 milli-arcseconds per
measurement (Eckart and Genzel 1996, 1997, Genzel et al. 1997). We used the
MPE-SHARP camera (Hofmann et al. 1993) on the 3.5m New Technology Telescope
(NTT) of the European Southern Observatory (ESO). SHARP contains a 256$^{{2}}
$ pixel NICMOS 3 detector. Each pixel projects to 25 or 50 milli-arcseconds
on the sky in order to (over-)sample the $\approx $ {}0.15'' FWHM
diffraction limited image of the NTT in K-band. The raw data for each data
set consist of several thousands of short exposure frames (0.3 to 0.5
seconds integration time). First we processed the data from  nights with
very good seeing (0.4 to 0.8'' ) in the standard manner (dead pixel
correction, sky subtraction, flat-fielding etc.). Next we co-added with the
simple-shift-and-add algorithm (SSA: for details see Christou 1991 and
Eckart et al. 1994). The individual short exposure frames typically contain
only a few bright speckles so that the SSA algorithm is well suited for our
purpose. The bright infrared sources IRS7 or IRS16NE serve as reference
sources. For the present study we analysed $\approx ${}82 independent data
sets from a total of 9 observing runs in 1992.25, 1992.65, 1993.65, 1994.27,
1995.6, 1996.25, 1996.43, 1997.54 and 1998.37. For the central arcsecond region
 around  SgrA* (the 'SgrA* cluster') we also analysed an additional
data set from 1999.42. Eckart and Genzel (1996,
1997) and Genzel et al. (1997 have previously analysed and discussed the
data until 1996.43. \newline
{\vskip0.21cm}

The diffraction limited core of the stellar SSA images contains up to 20
percent of the light. Determinations of the relative pixel offsets from
IRS16NE in raw SSA images or from diffraction limited maps after removing
the seeing halo give consistent results (see Eckart and Genzel 1996,
1997 for details). These `CLEANed' SSA maps produce similar results as other
data reductions (Knox-Thompson, triple correlation) but give much higher
dynamical range (see Eckart et al. 1994 for a detailed discusssion). This is
essential in the crowded Galactic centre region where the difference between
the brightest and faintest stars in our images is almost 10 magnitudes. We
solved for the relative offsets, rotation angle, and for the linear and
quadratic distortions between individual frames from an over-determined
system of non-linear equations for a reference list of relatively isolated
bright stars (Eckart and Genzel 1996, 1997). Our final near-infrared
reference frame is tied to an accuracy of $\pm $30 milli-arcsecs to the VLA
radio frame through 5 stars that show SiO and H$_{{\ 2}}$O maser activity
and are common to both wavelength bands (Menten et al. 1997). The resulting
combined systematic errors in our proper motion velocity estimates probably
are about 30 km/s. In the best cases these systematic effects now dominate
the error budget.

The new 1997 and 1998 data sets are in excellent agreement with the
extrapolation of the data we have published before and significantly improve
the uncertainties. As examples we show in {Fig.~(\ref{f:genzel1})} the
relative RA- and Dec-position offsets as a function of time between 1992 and
1998 for two selected stars. IRS16 C is a bright and isolated, HeI emission
line star (Krabbe et al. 1991, 1995). Its position vs. time diagram in {
Fig.~(\ref{f:genzel1})} is an example of the quality of the data on
bright isolated stars.
S1 is a faint star in the `SgrA* cluster' that is very close to SgrA* ($
\approx ${}0.1''). It shows the fastest proper motion ($\approx ${}1470
km/s) in the entire sample.{\vskip0.21cm}

\subsection{\bf 3D Spectroscopy}

We observed the Galactic centre with the MPE-3D near-infrared spectrometer
(Weitzel et al. 1996) in conjunction with the tip-tilt adaptive optics
module ROGUE (Thatte et al. 1995). 3D is a field imaging spectrometer which
obtains  spectra simultaneously for 256 spatial pixels covering a
square region of the sky (16x16 pixels). The fill factor is over 95\. For
further details of the instrument and data reduction we refer to Weitzel et
al. (1996). We observed the Galactic centre in March 1996 at the 2.2m
ESO-MPG telescope on La Silla, Chile. During the run the seeing on
the seeing monitor ranged between 0.3'' and 0.8''. The pixel scale was
0.3''. We observed the short-wavelength part of the K-band (1.9-2.2$\mu $m)
at $\lambda $/$\Delta \lambda $=2000, Nyquist sampled with two settings of a
piezo driven flat mirror. We covered the central $\approx ${}16''x10''
centered on SgrA* by an overlapping set of frames, each with a field of view
of 4.8''x4.8''. At each position we set up a sequence on-source (piezo step
1), off-source (piezo step 1), on-source (piezo step 2), off-source (piezo
step 2) etc. with an integration time per step of 200 seconds. Due to the
combined effects of seeing and pixel scale the resulting FWHM spatial
resolution of the final combined data set was 0.6''. We employed the
standard 3D data analysis package (based on GIPSY: van der Hulst et al.
1992). We performed wavelength calibration, sky subtraction, spectral and
spatial flat fielding, dead and hot pixel correction and division by a
reference stellar spectrum obtained during the observations. We corrected
for the effects of a spatially varying fringing or `channel' spectrum due to
interference in the saphire coating of the NICMOS 3 detector by applying
suitable flat fields from a set of flatfields at different settings of the
piezo mirror. Based on obervations of calibration lamps and OH sky lines
during the different observing nights the final velocity calibration is
accurate to $\pm 16km/s.${\vskip
0.21cm}

In the observed field we identified 21 emission line stars from continuum
subtracted images of the 2.058$\mu $m n=2 $^{{1}}$P- n=2 $^{{1}}$S and 2.11$
\mu $m n=4 $^{{3,1}}$S - n=3 $^{{3,1}}$P HeI lines and the 2.166$\mu $m
n=7-4 HI (Br$\gamma $) line. Most of the stars are identical with those
found in the $\approx ${}1'', R=1000 3D data set of Krabbe et al. (1995) and
Genzel et al. (1996) ( see also Blum, Sellgren and dePoy 1995 a,b and Haller
et al.1986) but the resolution, quality and nebular rejection is now much
superior. Three new stars were identified: (-2.1'', -4.1'') (13S SE),
(+1.6'', +0.3'') (16 CC) and (-8.3'', -5.7'') (all offsets are in RA and Dec
from SgrA*). We extracted from the data cube spectra of individual stars by
typically averaging 3 to 16 pixels on the star, for effective apertures
between 0.3'' and 1.2''. In most cases, we subtracted a suitable `off-star'
spectrum (scaled to the same aperture area) to eliminate the effect of local
nebular line emission. {Fig.~(\ref{f:genzel2a})} and {Fig.~(\ref{f:genzel2b})}
show the final spectra for 18 of the 21 stars.{\vskip0.21cm}

To determine stellar velocities we fitted Gaussians to the 2.058$\mu
$m and 2.11$\mu $m HeI lines and the 2.166$\mu $m Br$\gamma $ line.In
a few cases we also used in addition the 2.137/2.143$\mu $m MgII lines and the
2.189$\mu $m HeII line. A number of the stars clearly display P-Cyg
profiles in the 2.058$\mu $m HeI transition ({Fig.~(\ref{f:genzel2a})}
and {Fig.~(\ref{f:genzel2b})}) .  In these cases we fitted the profile
with a double Gaussian (absorption and emission). As the absorption
structure is well resolved an unambiguous emission line centroid
(assumed to be the stellar velocity) can thus be easily obtained. For
most stars we determined the final stellar velocities from averaging
the values obtained from 3 (or 4) lines. The agreement between the
fits to the different lines is generally good or even excellent.  We
list the final velocities in Table~1 and the insets of {Fig.~(\ref
{f:genzel2a})} and {Fig.~(\ref{f:genzel2b})}. The new velocity
determinations agree with those of Genzel et al. (1996) but the
uncertainties are typically half of those in our earlier work. The
best cases have an uncertainty of $ \pm $25 km/s.{\vskip0.21cm}

\begin{figure}
\centering \includegraphics[width=1.40\columnwidth,angle=90]{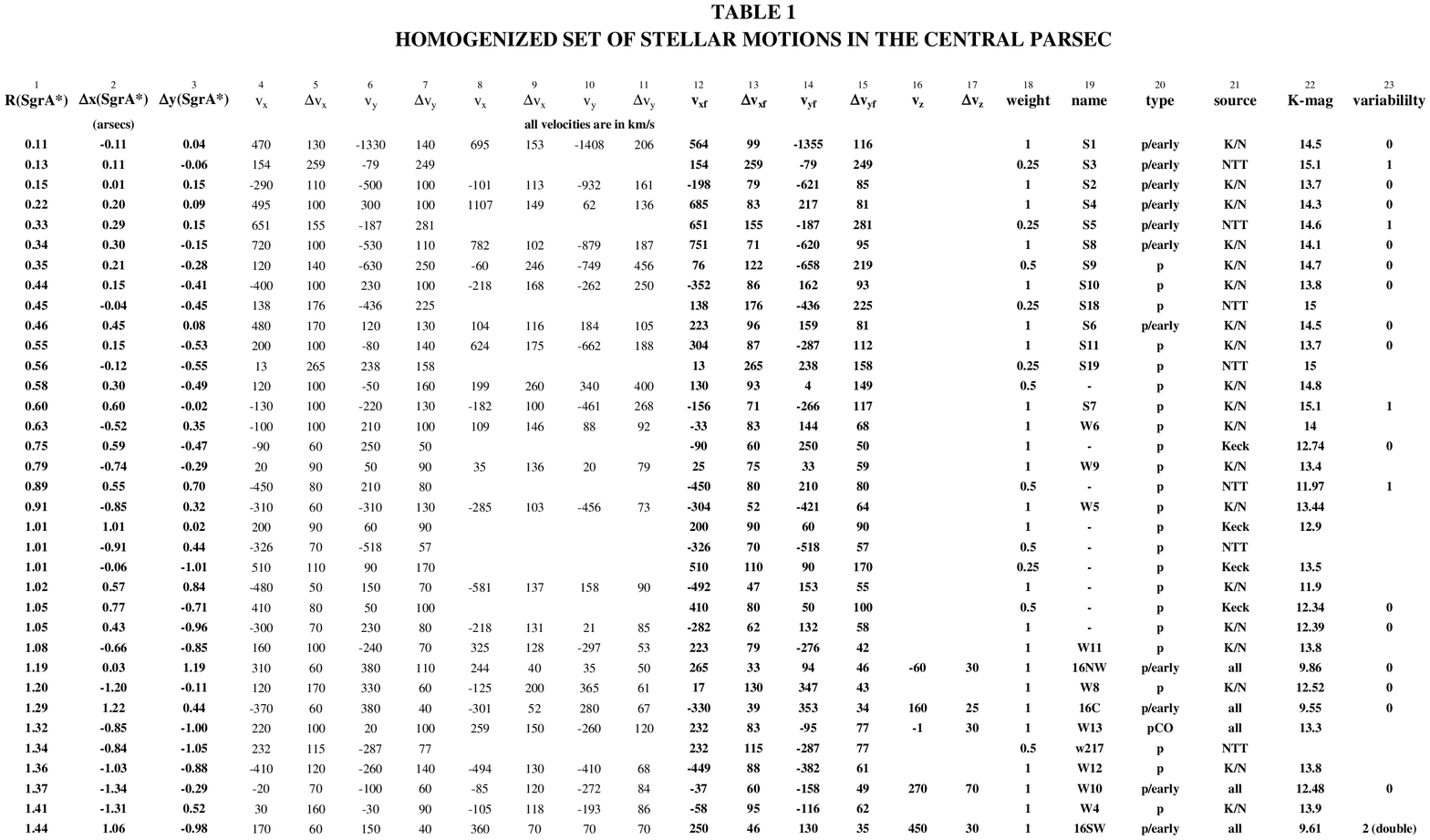}
\label{t:table1}
\end{figure}
\begin{figure}
\centering \includegraphics[width=1.40\columnwidth,angle=90]{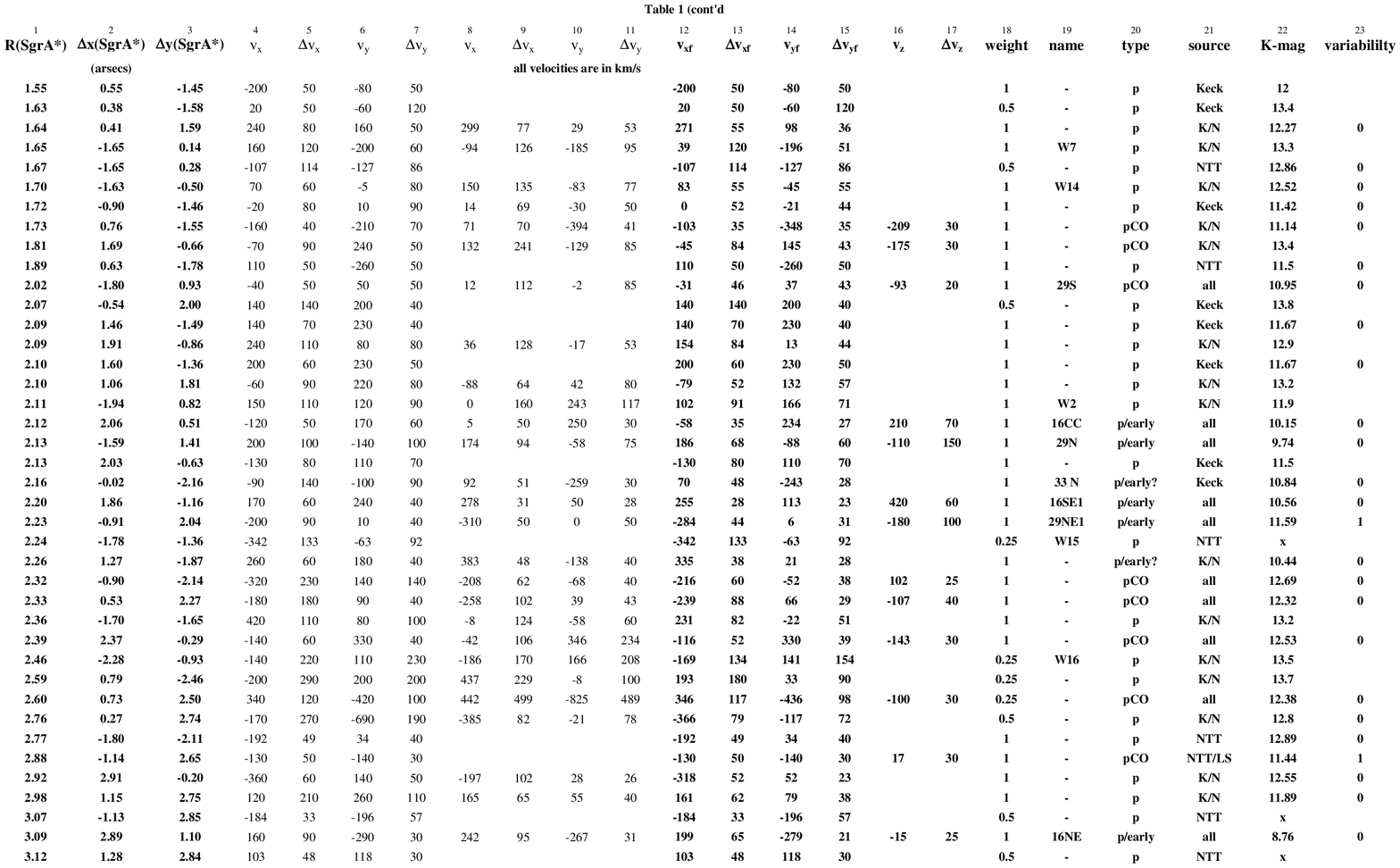}
\label{t:table1}
\end{figure}
\begin{figure}
\centering \includegraphics[width=1.40\columnwidth,angle=90]{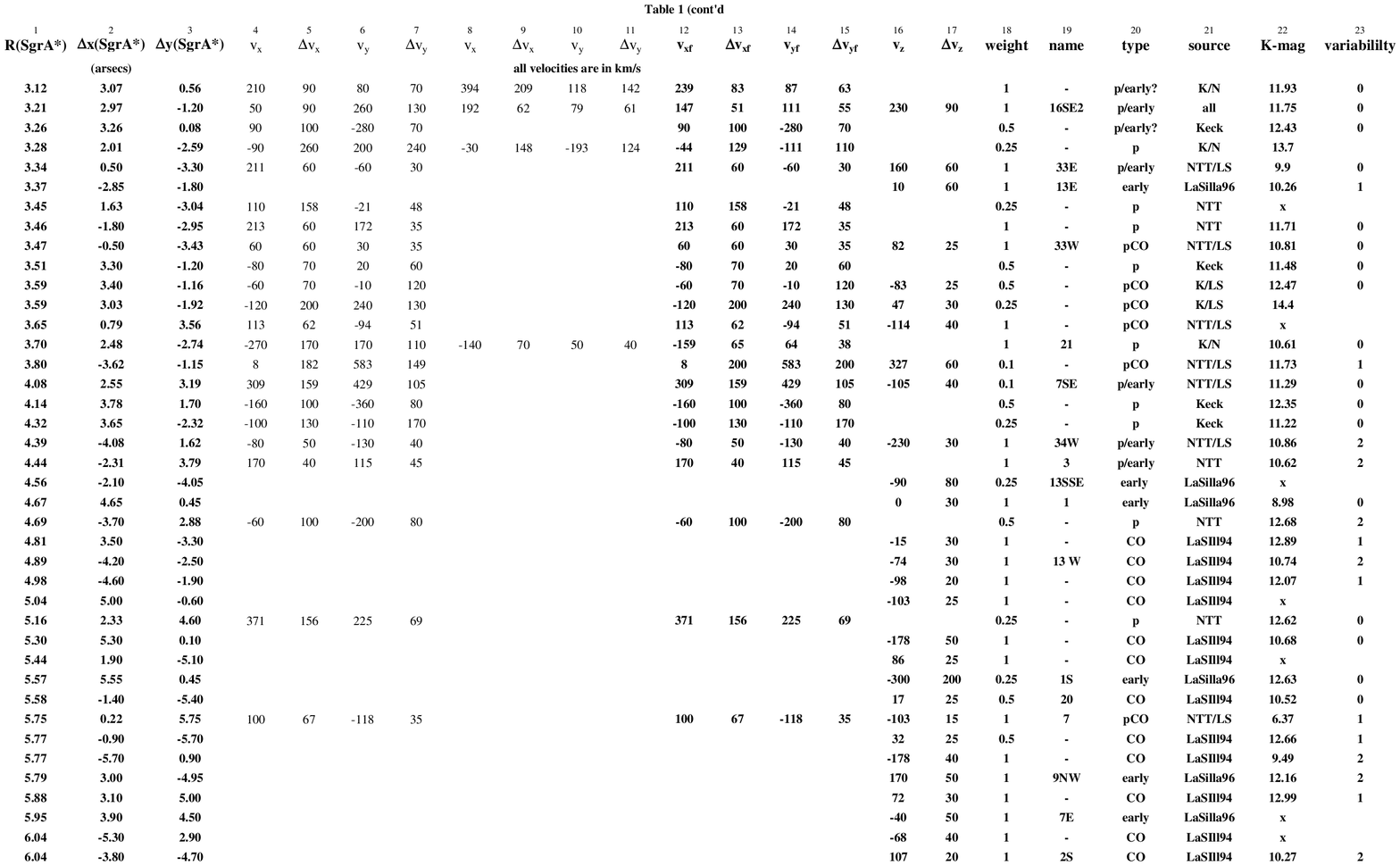}
\label{t:table1}
\end{figure}
\begin{figure}
\centering \includegraphics[width=1.40\columnwidth,angle=90]{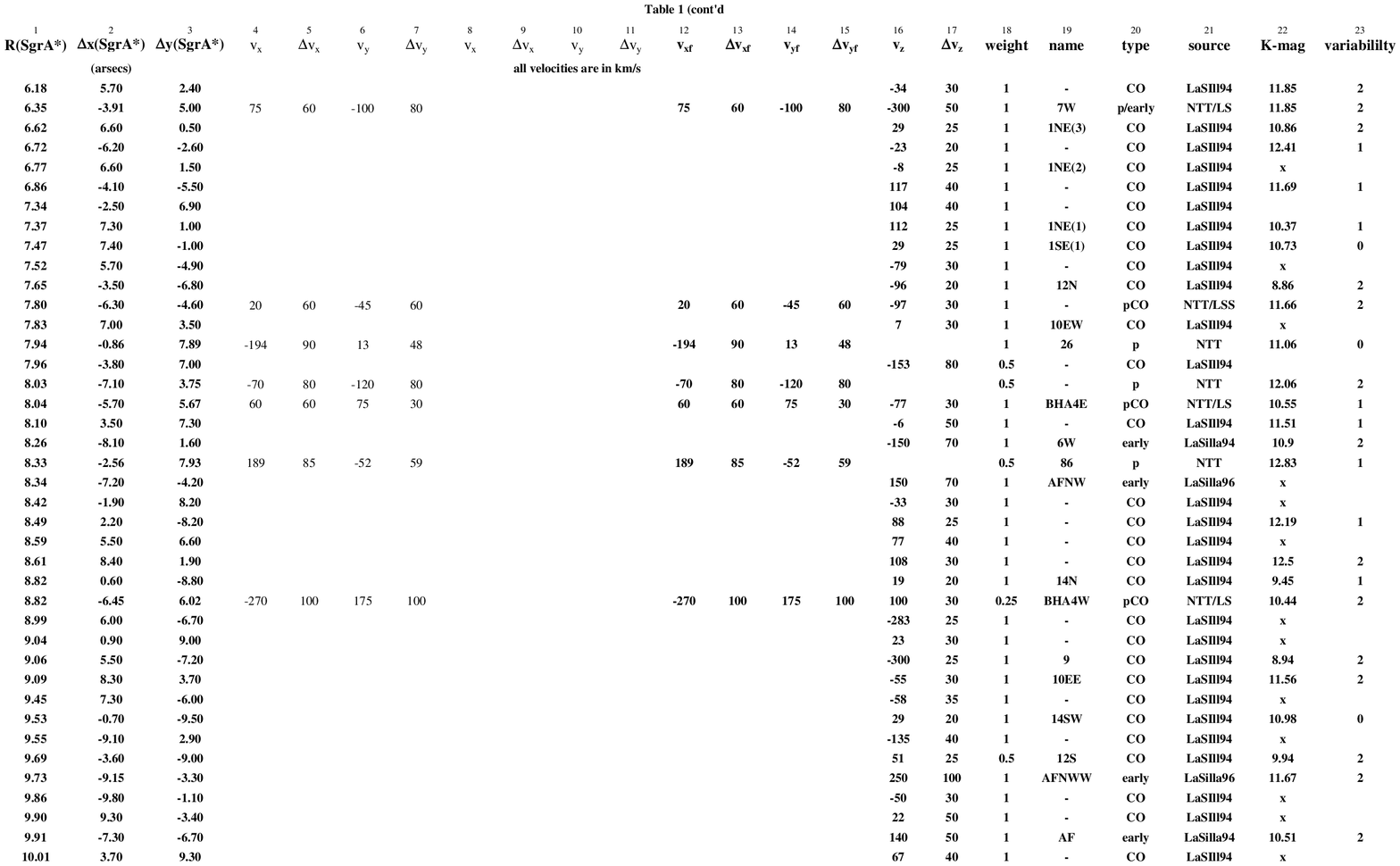}
\label{t:table1}
\end{figure}
\begin{figure}
\centering \includegraphics[width=1.40\columnwidth,angle=90]{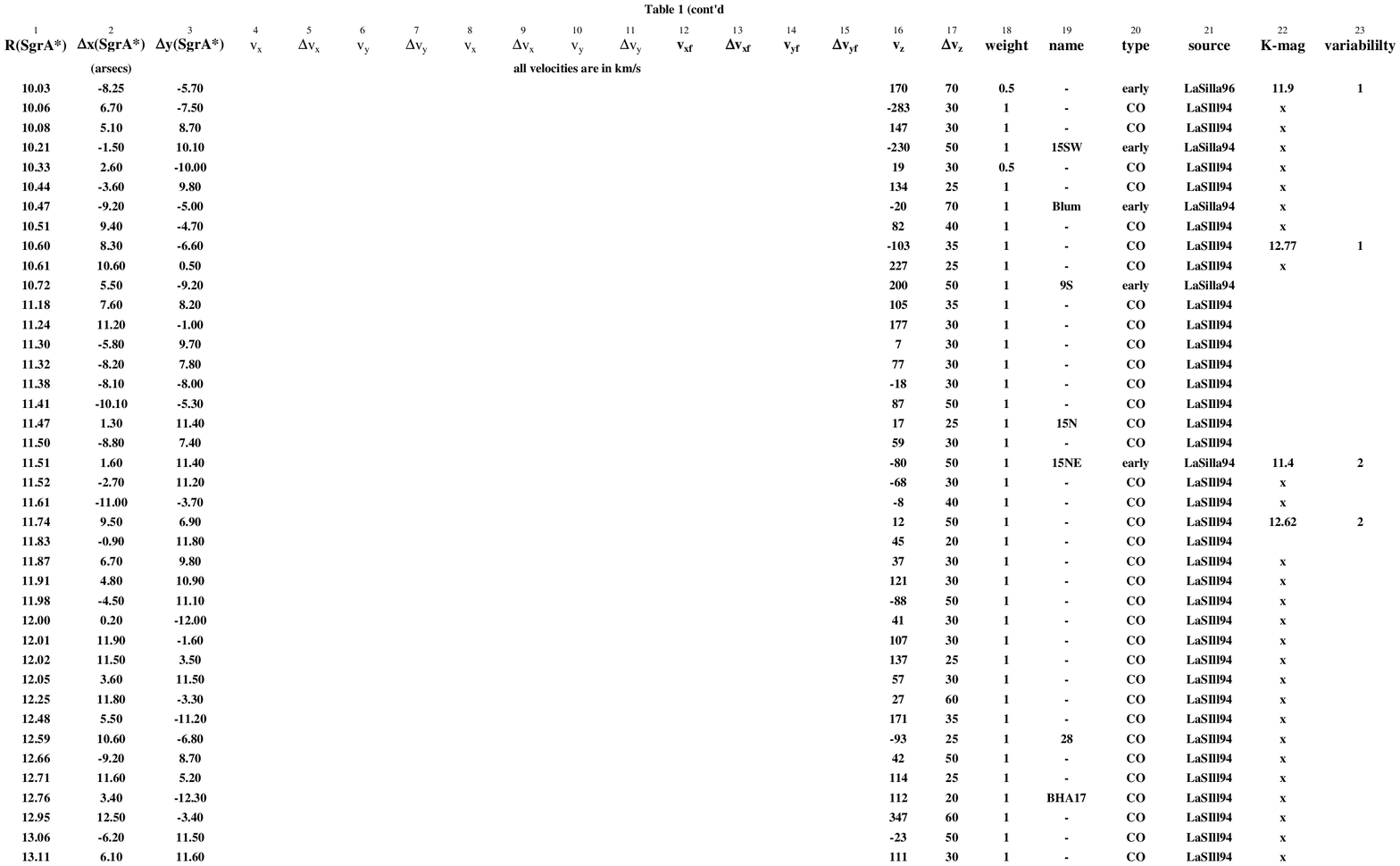}
\label{t:table1}
\end{figure}
\begin{figure}
\centering \includegraphics[width=1.40\columnwidth,angle=90]{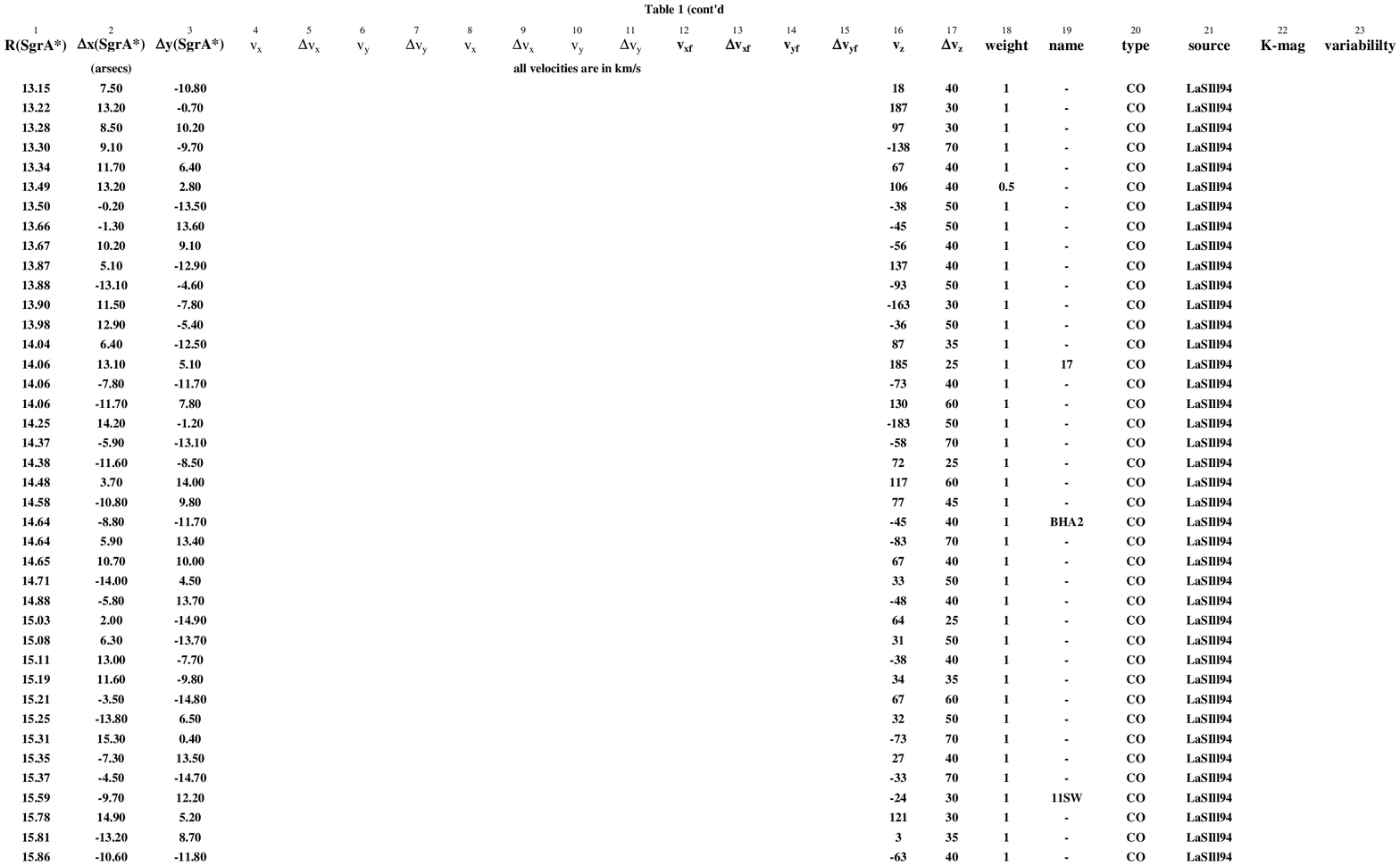}
\label{t:table1}
\end{figure}
\begin{figure}
\centering \includegraphics[width=1.40\columnwidth,angle=90]{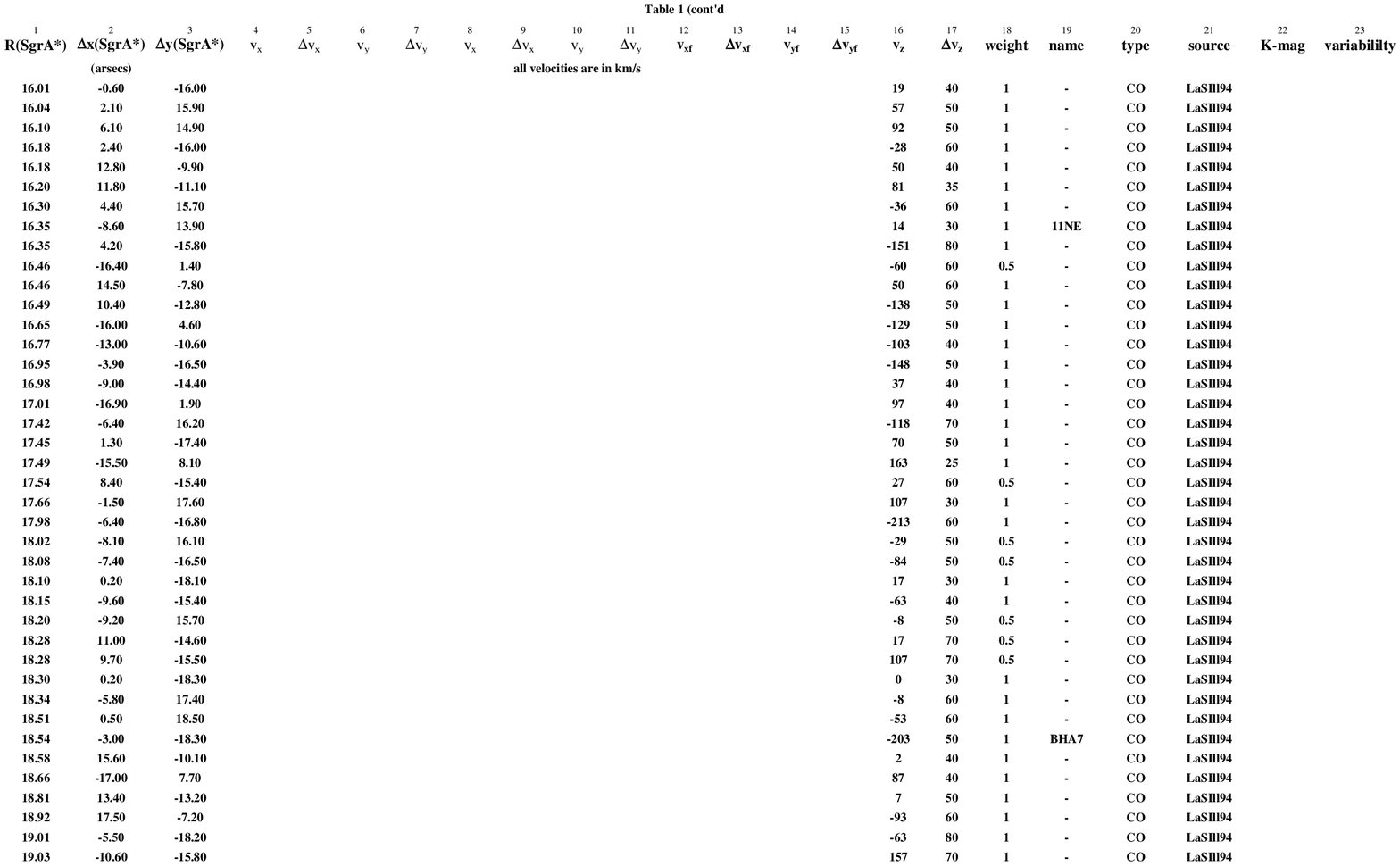}
\label{t:table1}
\end{figure}

\subsection{\bf A Homogenized Data Set}

To obtain a homogenized `best' data set of stellar velocities for further
analysis we combined the new 1992-1998 NTT proper motions and 2.2m
line-of-sight velocity data described in the last two paragraphs with the
1995-1997 Keck proper motion data of Ghez et al. (1998) and with other
relevant line-of-sight velocity data sets (see Genzel et al. 1996 and
references therein). Table~1 gives the results. The following explanations
and comments for Table~1 are in order.

\begin{figure}
\centering \includegraphics[width=1.00\columnwidth]{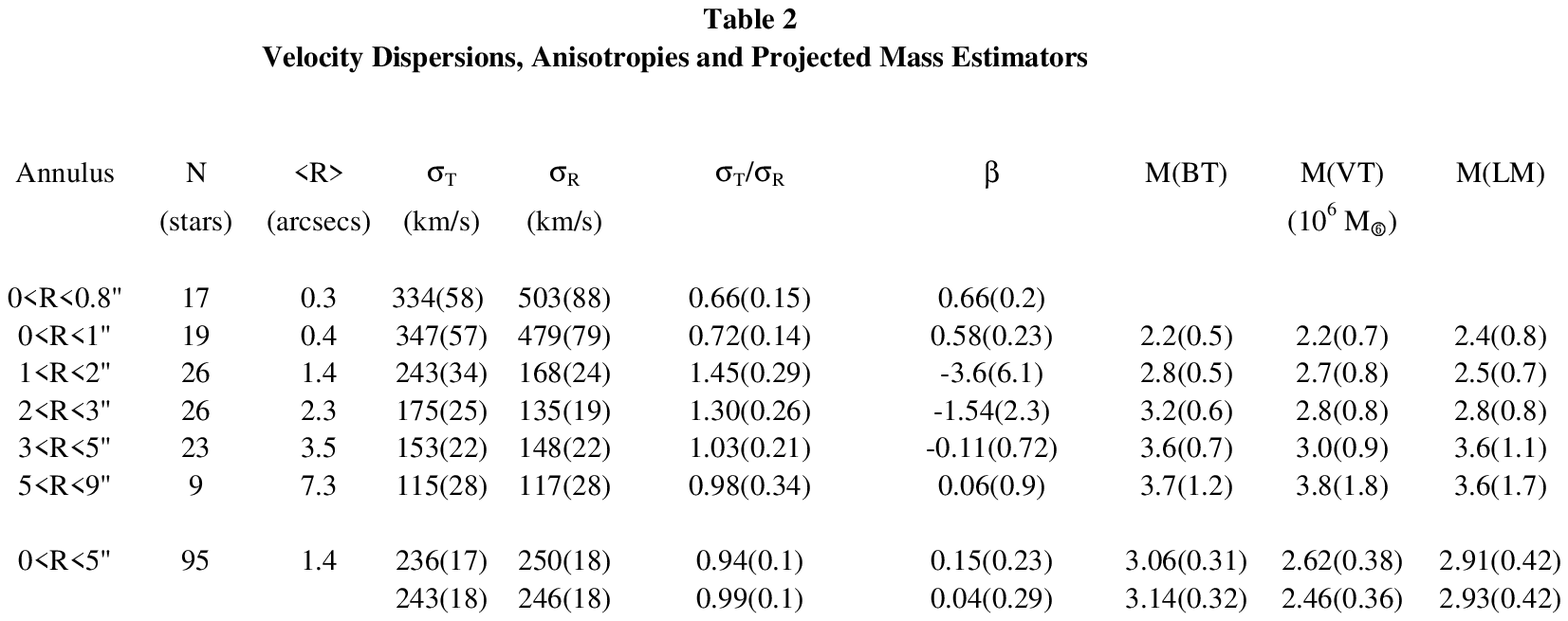}
\label{t:table1}
\end{figure}

$\bullet $ Columns 1 through 3 contain the projected separation R,
x(=RA)-offset and y (=Dec)-offset between star and SgrA* (epoch
1994/1995, all in arcseconds). As for the NTT, the Keck astrometry is
established with H$_{{2}}$O/SiO maser stars that are visible in both
wavelength bands and is accurate to $\pm $10 milli-arcseconds (see
Ghez et al. 1998, Menten et al.  1997).

$\bullet $ Columns 4 through 15 contain the x- and y-proper motions and
their respective 1$\sigma $ errors (km/s, for a Sun-Galactic centre distance
of $R_{{\sun}}=8.0 {{\rm kpc}}$). Whenever two measurements are available,
we first list the velocity from the Keck observations and then that from the
NTT observations. Columns 12 through 15 give the final combined proper
motions obtained from averaging motions from the two sets (if available)
with 1/$\sigma ^{{2}}$ weighting. The agreement between Keck and NTT data
sets generally is very good (see discussion by Ghez et al. 1998) and is in
accordance with the measurement uncertainties. We thus assume that the final
measurement error of the combined set is given by 1/ $\surd $ (1/$\sigma _{{
1 }}^{{2}}$ + 1/$\sigma _{{2}}^{{2}}$). We have found, however, from a
comparison of the two data sets that stars with $>$200 km/s velocity
uncertainty at $R\geq 1^{\prime\prime}$ (and $>$400 km/s at $R\leq
1^{\prime\prime}$) are highly unreliable. We therefore have eliminated such
stars from the final set.

$\bullet$ Columns 16 and 17 give the line of sight velocity and its 1$\sigma$
error.

$\bullet$ Column 18 assigns a weight to each data point, based on its
reliability, the velocity errors and the agreement between different data
sets. The weight is approximately proportional to 1/error$^{{2}}$ as
appropriate for white noise but `quantized' in order to not place too much
weight on the few data points with the smallest statistical errors. While
this weighting scheme is subjective it is in our opinion a fair
representation of the quality of the different data points in the presence
of significant systematic errors. We have also tried another, more formal
weighting scheme. Here we have assigned the weight $w = 1/ (1 + ({{\rm 
error}}/ \sigma_{{0}})^{{2}}$), where `error' is the x/y-averaged proper
motion velocity uncertainty. $\sigma_{{0}}=250 R^{{-0.5}}$ is a measure of
the sample dispersion at R. This weighting scheme also gives essentially the
same weights for different data points as long as their individual errors
are much smaller than the sample velocity dispersion. For large errors the
weight scales as 1/error$^{{2}}$ as for white noise. We have applied both
weighting schemes in the various estimates discussed in the text. The
results are basically identical.

$\bullet$ Column 19 lists the `popular' name of the star.

$\bullet$ Column 20 lists an identification; `p' stands for a star with a
measured proper motion, `early' denotes that the star is a young early type
star (e.g. HeI/HI emission lines), and `CO' denotes that the star is a late
type star.

$\bullet$ Column 21 lists the source(s) of measurement for the specific data
point (Keck (K) and NTT (N) for proper motions, LaSilla (LS), or `all').

$\bullet $ Column 22 lists the K-magnitude of the star, and column 23 makes
a statement on its variability. If available, we used the
K-magnitudes from the comprehensive variability study of Ott et al. (1999),
otherwise we list the magnitudes of Ghez et al. (1998), corrected for $
\delta $m$_{{K}}$=-0.4 to account for a small calibration offset between the
Ott et al. and Ghez et al. sets. If the value in the last column is 0, the
star was not or not significantly variable in the 1992 to 1998 monitoring
campaign of Ott et al. A value of 1 indicates that the star showed a
statistically significant but weak variability. A value of 2 indicates that
the star was strongly variable in the Ott et al. observations.{\vskip0.21cm}

\section{\bf Kinematics of the Galactic Centre Star Cluster}

\begin{figure}
\centering \includegraphics[width=0.75\columnwidth,clip=true]{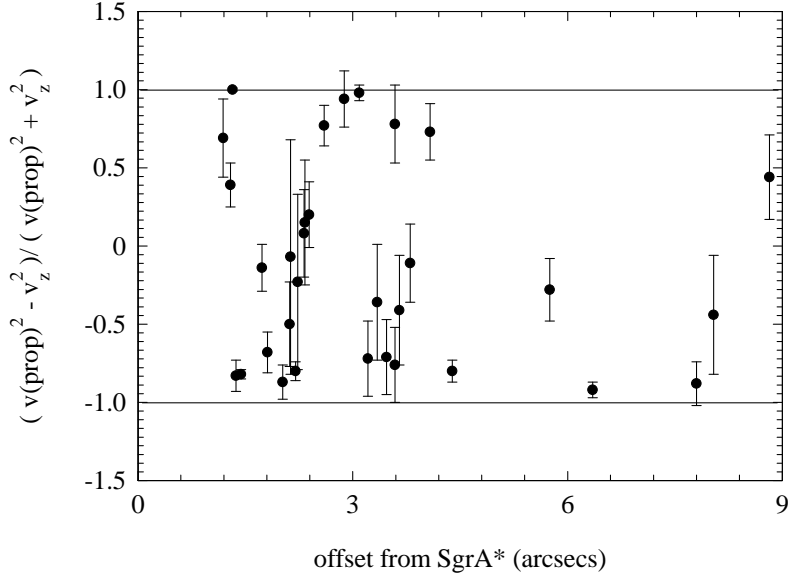}
\vskip -3cm
\caption{ The anisotropy parameter $\protect\gamma _{{pz}}= (v_{{prop }
}^{{2}} - v_{{z}}^{{2}})  / (v_{{prop}}^{{2}} +  v_{{z}}^{{2}})$  for all 32
stars with all three velocity components measured. Error bars are determined
through error propagation from the velocity uncertainties in
Table~1. }
\label{f:genzel3}
\end{figure}
\par

\begin{figure}
\centering \includegraphics[width=0.95\columnwidth,clip=true]{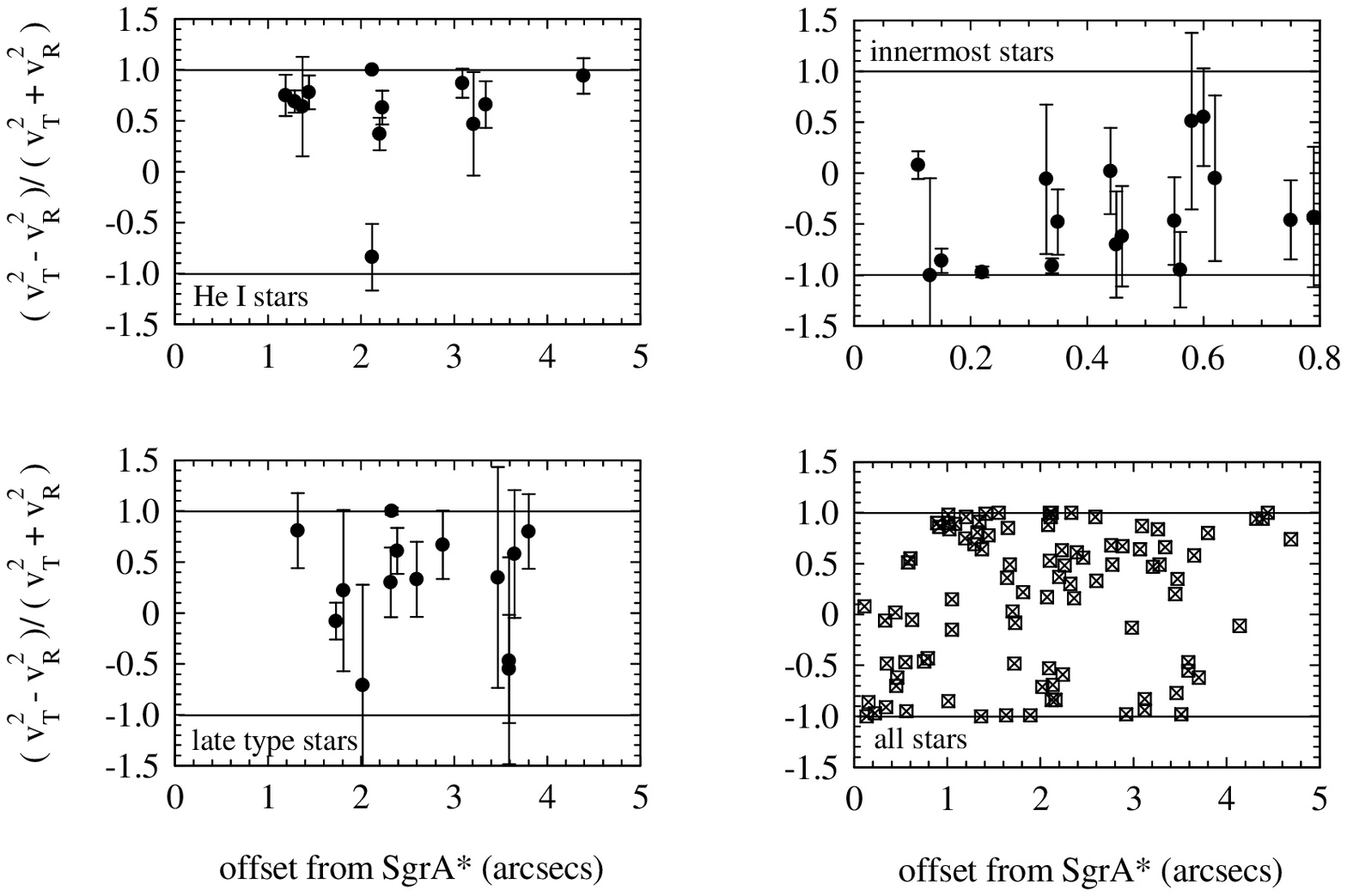}
 \vskip -3.cm

\caption{ The anisotropy parameter $\protect\gamma _{{TR}}= (v_{{T}}^{{\
2}} - v_{{R}}^{{2}}) / (v_{{T}}^{{2}} + v_{{R}}^{{2}}$) for the 12 HeI stars
within $R\leq 5``$ (top left), for the 14 late type stars within $R\leq 5``$
(bottom left), for  all stars inside 5`` with  proper motions (bottom right)
and for stars within  $R\leq  0.8``$ of  SgrA*  (top right). Error  bars are
determined   through error propagation   from  the velocity uncertainties in
Table~1. }
\label{f:genzel4}
\end{figure}

The  velocity determinations in Table~1  are significantly improved over our
earlier work  and  over  Ghez et  al. (1998).   Many light-of-sight and  sky
velocities  now have  errors less than  50 $\kms$,   with the best  velocity
determinations ($\pm $20 to  25 km/s) mainly  limited by  systematic effects
(e.g.  establishing reference frame  from  the  moving stars themselves  and
removing distortions  in the imaging),   rather than by  statistical  errors
(positional accuracy  of  stellar positions). Of   the  $2\times 104$ proper
motions  in Table~1, 48  (23\%) are determined to   4$\sigma $ or better. 5
proper motions are determined at  the $\geq $10$\sigma $  level. Of the  227
line-of-sight velocities 38  (17\%) are determined to  4$\sigma $ or better.
For 14 (of  29) HeI emission line  stars and for 18 late  type stars  we now
have determinations of all  three  velocity components. With this   improved
data set it  is  now possible to  investigate  in more detail  the kinematic
parameters of  individual  stars and/or small groups  of  stars. In order to
remove as well as possible  measurement and calibration  bias and zero point
offsets, we  subtract  in all of our  calculations  below for each  velocity
measurement i the mean velocity of the sample  $\langle v \rangle $ linearly
and the  velocity uncertainty  error (v$_{{i}}$)  in squares when  computing
velocity dispersions etc.,
\begin{equation}
(v_{{i}}^{{2}})_{{\rm intrinsic}}=(\{v_{{i}}-\langle v\rangle \}^{{2}})_{ 
{\rm measured}}-error^{{2}}(v_{{i}}).  \label{e:(1)}
\end{equation}

\subsection{ Tests for Anisotropy}

As proposed by Eckart and Genzel (1996, 1997), a first simple (but
coarse) test for anisotropy in the data (and/or a Sun-Galactic centre
distance significantly different from $R_{{\sun}}=8 {{\rm kpc}}$) is
to compare the sky and line-of-sight velocities of individual
stars. {Fig.~(\ref{f:genzel3}) } is a plot of $\gamma
_{{pz}}$=(v$_{{prop}}^{{2}}$ - v$_{{z}}^{{2} }$)/ (v$_{{prop}}^{{2}}$
+ v$_{{z}}^{{2}}$) as a function of projected separation R for the 32
stars with three measured velocities. Here v$_{{prop}}$ is the root
mean square of the x- and y-sky motions, $v_{{z}}$ is the
line-of-sight motion. In this plot stars with $\gamma _{{pz}}=-1$ have
v$_{ {z}} \gg v_{{prop}}$ stars with $\gamma _{{pz}}$=+1 have
$v_{{z}}\ll v_{{prop}}$. {Fig.~(\ref{f:genzel3})} shows no obvious
sign for such an anisotropy. This is probably not surprising as the
line-of-sight and sky velocities both contain linear combinations of
the intrinsic radial and tangential components of the velocity
ellipsoid. The result in {Fig.~(\ref {f:genzel3})} that the sample
expectation value for the proper motion velocity dispersion is the
same (within statistical uncertainties) as the line-of-sight
dispersion, $<v_{prop}^{2}>\simeq <v_{z}^{2}>$, is consistent with the
assumption that we are observing a spherically symmetric cluster.  The
virial theorem guarantees that this results holds independent of
internal anisotropy (see equations {~(\ref{e:(5a)})} {~(\ref{e:(5b)})}
{~(\ref{e:(5c)})}) .{\vskip0.21cm}

To investigate intrinsic kinematic anisotropies it is therefore necessary to
explicitly decompose the observed motions into projections of the intrinsic
velocity components. Assuming that the velocity ellipsoid of a selected
(sub-) sample of stars separates in spherical coordinates and denoting the
components of velocity dispersion parallel and perpendicular/tangential to
the radius vector {\bf {\it r \/}} as $\sigma _{{r}}$ and $\sigma _{{t}} 
$, the line-of-sight component $\hat{\sigma}_{{z}}(R,z)$ is then given by 
\begin{equation}
\hat{\sigma}_{{z}}^{{2}}=\sigma _{{r}}^{{2}}(r)\cos ^{{2}}\theta
+\sigma _{{t}}^{{2}}(r)\sin ^{{2}}\theta ,  \label{e:(2a)}
\end{equation}
where $\cos \theta ={{{\bf {r}}}}\cdot {{{\bf {z}}\/}}/r$ and ${} 
{\bf z\/}$ is the unit vector along the line-of-sight. The components of the
velocity dispersion parallel (R) and perpendicular (T) to the projected
radius vector on the sky {\bf {\it R\/}} are given by 
\begin{equation}
\hat{\sigma}_{{R}}^{{2}}=\sigma _{{r}}^{{2}}(r)\sin ^{{2}}\theta
+\sigma _{{t}}^{{2}}(r)\cos ^{{2}}\theta ,
\end{equation}
and 
\begin{equation}
\hat{\sigma}_{{T}}^{{2}}=\sigma _{{t}}^{{2}}(r)\quad .
\label{e:(2c)}
\end{equation}
\noindent Given the spatial density distribution $n(r)$ of the selected
sample of stars (assumed to be spherically symmetric) the line-of-sight
averaged, density weighted value of the projected radial velocity dispersion
of the sample at $R$, $\sigma _{{R}}$, can then be computed from the
relationship 
\begin{equation}
\Sigma (R)\sigma _{{R}}^{{2}}(R)=\int_{-\infty }^{\infty }n(z)(\sigma_{{
r}}^{{2}}(r)\sin^{{2}}\theta +\sigma_{{t}}^{{2}}(r)\cos^{{2}}
\theta ){{\rm d}{z}}=2\int_{R}^{\infty } [\sigma_{{r}}^{{2}}(r)(R/r)^{{2}
}+\sigma_{{t}}^{{2}}(r)(1-(R/r)^{{2}}] \frac{n(r)r{{\rm d}{r}}} {(r^{{
2}}-R^{{2}})^{{1/2}}}\quad  \label{e:(3)}
\end{equation}
where $\Sigma (R)$ is the stellar surface density at R, 
\begin{equation}
\Sigma (R)=2\int_{R}^{\infty }\frac{n(r)r{{\rm d}{r}}}{(r^{{2}}-R^{{2}
})^{{1/2}}}\quad  \label{e:(4)}
\end{equation}
Similar equations hold for $\sigma _{{T}}^{{2}}(R)$ and $\sigma _{{z}
}^{{2}}(R)$. The global expectation value of the projected radial
dispersion is given by 
\begin{equation}
\langle \sigma _{{R}}^{{2}}\rangle =(2\pi /N)\int_{0}^{\infty
}\int_{0}^{\pi }n(r)r^{{2}}(\sigma _{{r}}^{{2}}(r)\sin ^{{2}}\theta
+\sigma _{{t}}^{{2}}(r)\cos ^{{2}}\theta )\sin \theta {{\rm d}{\theta }
}{{\rm d}{r}}=2/3\langle \sigma _{{r}}^{{2}}\rangle +1/3\langle \sigma _{
{t}}^{{2}}\rangle \quad ,  \label{e:(5a)}
\end{equation}
where N is the number of stars in the selected sample. Likewise one finds 
\begin{equation}
\langle \sigma _{{z}}^{{2}}\rangle =(2\pi /N)\int_{0}^{\infty
}\int_{0}^{\pi }n(r)r^{{2}}(\sigma _{{r}}^{{2}}(r)\cos ^{{2}}\theta
+\sigma _{{t}}^{{2}}(r)\sin ^{{2}}\theta )\sin \theta {{\rm d}{\theta }
}{{\rm d}{r}}=1/3\langle \sigma _{{r}}^{{2}}\rangle +2/3\langle \sigma _{
{t}}^{{2}}\rangle ,  \label{e:(5b)}
\end{equation}
and 
\begin{equation}
\langle \sigma _{{T}}^{{2}}\rangle =\langle \sigma _{{t}}^{{2}
}\rangle  \label{e:(5c)}
\end{equation}
(Leonard \& Merritt 1989). Deviations of the velocity ellipsoid from
isotropy are commonly expressed in terms of the anisotropy parameter $\beta
=1-\sigma _{{t}}^{{2}}/\sigma _{{r}}^{{2}}$. Its globally averaged
value is given by 
\begin{equation}
\langle \beta \rangle =1-\langle \sigma _{{t}}^{{2}}\rangle /\langle
\sigma _{{r}}^{{2}}\rangle =3(\langle \sigma _{{R}}^{{2}}\rangle
-\langle \sigma _{{T}}^{{2}}\rangle )/(3\langle \sigma _{{R}}^{{2}
}\rangle -\langle \sigma _{{T}}^{{2}}\rangle )\quad .  \label{e:(6)}
\end{equation}
An isotropic cluster ($\langle \beta \rangle $=0) has $\langle \sigma _{{r}
}\rangle $=$\langle \sigma _{{t}}\rangle $ and $\langle \sigma _{{R}
}\rangle $=$\langle \sigma _{{T}}\rangle $. A cluster with only radial
orbits ($\langle \beta \rangle $=1) has $\langle \sigma _{{t}}\rangle $=$
\langle \sigma _{{T}}\rangle $=0 or $\langle \sigma _{{R}}\rangle \gg
\langle \sigma _{{T}}\rangle $. A cluster with only tangential orbits ($
\langle \beta \rangle $ = -$\infty $) has $\langle \sigma _{{r}}\rangle $
=0 and $\langle \sigma _{{R}}\rangle $= $\langle \sigma _{{T}}\rangle $/$
\surd $3. Thus radial anisotropy is easier to see in the proper motions than
tangential anisotropy. Table~2 gives the values of $\langle \sigma _{{R}
}\rangle $, $\langle \sigma _{{T}}\rangle $, $\langle \sigma _{{R}
}\rangle $/$\langle \sigma _{{T}}\rangle $ and $\langle \beta \rangle $,
computed for all stars with proper motions and for different ranges of
projected radii from SgrA*. Errors in these quantities are derived  from
statistics and error propagation. Below we use Monte Carlo simulations to
investigate the uncertainties in the derived anisotropy parameters more
throroughly. The proper motions of the entire sample of stars as well as the
stars in the range R$\geq $3'' are consistent with isotropy. At R=1 to 3''
there is a (marginal) trend for the stars to be more on tangential orbits.
In the central arcsecond the stars on average appear to be on radial orbits.
The statistical significance of this departure from isotropy for the 17
stars at R$\leq $0.8'' from SgrA* appears to be 3.3$\sigma $ in terms of
propagated errors for $\beta$ (Table~2); however, the Monte Carlo
simulations of Section 4.3 below show that the distribution of $\beta$ is
very broad and the isotropic $\beta=0$ is still within somewhat more than 1$
\sigma $ equivalent for such a small sample. 
Excluding the faint stars in the SgrA* cluster at R$\leq $
1'' the remaining proper motions between R=1'' and 5'' deviate from isotropy
in the direction of tangential orbits at the 2$\sigma $ level. 
\vskip0.21cm

A second and more sensitive test for anisotropy is a comparison of the
projected radial (v$_{{R}}$) and projected tangential (v$_{{T}}$)
components of the sky velocities of individual stars. {Fig.~(\ref{f:genzel4})
} gives plots of $\gamma_{TR}=(v_{{T}}^{{2}} - v_{{R}}^{{2}})/ (v_{{
T}}^{{2}} + v_{{R}}^{{2}})$ for different selections of our data.
Considering all 104 proper motions, {Fig.~(\ref{f:genzel4})} (bottom right)
indicates a fairly even distribution of $\gamma _{{TR}}$s, without obvious
overall bias indicating anisotropy, perhaps a slight predominance of
tangential orbits (compare {Fig.~(\ref{f:genzel8})}). The same is true if
only the late type stars of the proper motion sample are considered ({Fig.~(
\ref{f:genzel4})}, bottom left).

A different and fairly clearcut picture emerges when one considers the (much
younger) early type stars. {Fig.~(\ref{f:genzel4})} (top left) clearly
indicates that with one exception all bright HeI emission line stars within
R=5'' are on projected tangential orbits ($\gamma _{{TR}}$$\approx ${}+1)
\noindent and therefore (see the discussion after equation {~(\ref{e:(6)})} 
), largely on true tangential or circular orbits. In contrast, more than
half of the faint (m$_{{K}}$$\approx ${}13 to 16) stars within 1'' of
SgrA* (SgrA* cluster) are predominantly on radial ($\gamma _{{TR}}\leq 
$ 0) orbits (top right panel of {Fig.~(\ref{f:genzel4})}). {\bf {\it We
conclude that the early type stars in our proper motion set do show
significant anisotropy.\/}}. We will show below that the main cause
of the tangential anisotropy is a global rotation of the
early type stars.{\vskip0.21cm}

\subsection{ Orbits for the innermost stars}

\begin{figure}
\centerline{\psfig{width=1.20\columnwidth,clip=true,file=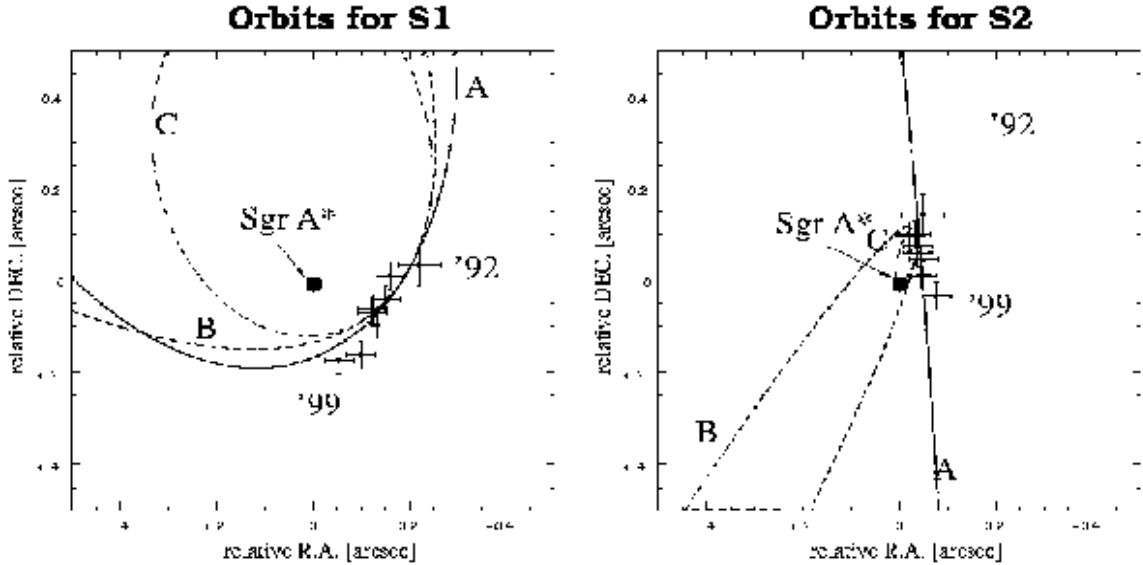}}
\caption{ Possible orbits for S1 and S2, the stars closest to SgrA*. The
positions of S1(left) and S2 (right) between 1992 and 1999 and their
uncertainties are shown as crosses. Three possible bound orbits in the
potential of a  2 to $3\times 10^{{6}} M_{{\sun}}$ point mass are shown.
Orbit A (continuous curve) has the largest possible separation from SgrA*
and orbit B (short dashes) the largest line of sight velocity. The
assumption that S1 and S2 are in the same plane of the sky as the central
mass and have no line of sight velocity results in the orbit with the
largest curvature (orbit C, dash-dotted). }
\label{f:genzel5}
\end{figure}

We have also modeled the orbits of several of the individual fast
moving stars in the SgrA* cluster. As an example we plot in
{Fig.~(\ref{f:genzel5})} the measured 1992-1999 NTT positions of S1
and S2 with respect to SgrA*, along with the projection of a few
possible trajectories. The three plotted orbits represent extreme
choices of the orbital parameters in the potential of a central
compact mass. For orbit A we assumed the largest possible current
separation from SgrA* for bound orbits with $v_z=0$. For orbit B we
took the largest line of sight velocity at $z=0$ under the boundary
condition that S1 and S2 are still bound to SgrA*. The assumption that
S1/S2 are in the same plane of the sky as the central mass and have no
line of sight velocity results in the orbit with the largest curvature
(orbit C).  Although no unique orbit can yet be determined from the
data, our analysis shows that most of the high velocity stars in the
SgrA* cluster can be bound to a central mass of $~3\times 10^{{6}}
M_{{\sun }}$ with a distribution of line of sight positions and
velocities that is consistent with the projected dimensions and
velocity dispersion of the cluster{\bf {\it .  Orbits with radii of
curvature comparable to their projected radii from SgrA* (orbit C) can
already be excluded for these stars\/}}. S1, S2 and several other fast
moving stars around SgrA* must be on plunging (radial) orbits or on
very elliptical/parabolic orbits with semi-axes much greater than the
current projected separations from SgrA*, as already indicated by the
analysis of the velocities in the preceding paragraph. The stellar
orbits in the central cluster will not be simple closed
ellipses. Especially for orbits with large eccentricities,
node-rotation due to the non-Keplerian potential in the extended
stellar cluster as well as relativistic periastron rotations will make
the trajectories for the individual stars `rosette'-like. A more
detailed analysis of orbits has to await longer time baselines for the
proper motion measurements and the detection of orbit curvature (=
acceleration), as well as measurements of the line of sight velocities
of the central stars.{\vskip0.21cm}

\subsection{\bf Anisotropy and Relaxation Time}

These deviations from isotropy for the early-type stars are consistent with
their young ages as compared to the relaxation time. Within the central
stellar core, the two-body relaxation time for a star of mass m$_{{10}}$
(in units of 10 M$_{{\odot }}$) is given by 
\begin{equation}
t_{{r}}(m)=10^{{7.58}}\sigma _{{100}}^{{3}}/^{{}}(\rho _{{6.6}}m_{{
10}}\{lnN_{{\ast }}/13\})\quad ({\rm yrs}).  \label{e:(7)}
\end{equation}
Here $\rho _{{6.6}}$ is the density of the nuclear star cluster (in units
of $4\times 10^{{6}} M_{\sun } {{\rm pc}}^{{-3}}$) and $\sigma _{{100
}}$ is the velocity dispersion in units of 100 km/s within the core radius
\footnote[1]{
The core radius is defined here as the radius where the stellar density has
fallen to half its central value.\newline
} of 0.38 (+0.25,-0.15) pc (Genzel et al. 1996). N$_{{\ast }}$ is the
number of stars in the core ($\approx {}3 to 5\times 10^{{5}}$). The
life-time of the upper main sequence phase scales approximately as $t_{{ms}
} \approx {}10^{{7.2}} m_{{10}}^{{-1.9}}{{\rm yr}}$ and the duration
of the red-/blue-giant or supergiant phases is typically 10 to 30\% of $t_{{
ms}}$ \noindent (e.g. Meynet et al. 1994). The ratio of relaxation time to
stellar life time thus is 
\begin{equation}
\chi =t_{{r}}/t_{{ms}}\approx {}2.4m_{{10}}^{{0.9}}.  \label{e:(8)}
\end{equation}
A number of authors have shown have shown that the HeI emission line stars
are high mass (30 to 120 M$_{\sun }$), post main sequence blue supergiants
(Allen et al. 1990, Krabbe et al. 1991, 1995, Najarro et al. 1994, 1997,
Libonate et al. 1995, Blum, Sellgren and dePoy 1995 a,b, Tamblyn et al.
1996, Ott et al. 1999). Their ages range between $t\approx {}2 {{\rm and 
}} 9\times 10^{{6}}$ yrs (Najarro et al. 1994, 1997, Krabbe et al. 1995).
The massive stars are probably the `tip of the iceberg' of a component of
young stars of total mass $\approx ${}10$^{{4}}$ M$_{\sun }$ that were
formed a few million years ago in an extended starburst episode (Krabbe et
al. 1995). The massive stars are somewhat older than their main sequence
age. Their main sequence life time is much greater than the dynamical time
scale ($t_{ms}\gg t_{{\rm dyn}}\simeq 10^{3}{\rm yr}$) but they have not had
time to dynamically relax through multiple interactions with other stars ($
\chi \gg 1$). Their present kinematic properties thus reflect the initial
conditions with which they were born and the starburst must have been
triggered near their present orbits. The situation is different for the
observed late type stars. For M-giants of mass 1.5 to 3 M$_{\sun }$ (and
ages $\geq $10$^{{9} }$ years) and for luminous asymptotic giant branch
(AGB) stars of mass 3 to 8 M$_{\sun }$ (and ages $\geq $10$^{{8}}$ years), 
$\chi $ is comparable to or smaller than unity. Such stars should have had
sufficient time to be scattered and relax in the central potential.{\vskip
0.21cm}

\section{ Monte Carlo Simulations}

\label{s:mts}

Because the velocity measurement errors are often large and the number of
measured velocities is still relatively small, we need to investigate the
expected errors in the velocity anisotropy in more detail to get a more
quantitative estimate whether the observed anisotropy is statistically
significant. Therefore we now describe theoretical 'measurements' on Monte
Carlo star clusters with comparable numbers of stars. The next subsection
describes the models from which the artificial data are drawn.

\subsection{Anisotropic distribution functions}

We construct some simple anisotropic, scale-free spherical distribution
functions $f(\varepsilon ,h)$ for stars with specific energy $\varepsilon $
and specific angular momentum $h$ in a potential $\psi $. These are computed
from the formula (for a derivation and its generalization to non integer
index $n$, see, e.g., Pichon \& Gerhard, in preparation): 
\begin{equation}
f(\varepsilon ,h)=(-\varepsilon )^{n+{\frac{1}{2}}}{\frac{\sqrt{2}\
}{{4\pi \sqrt{\pi }}\,{\Gamma }({\frac{3}{2}}
+n)}}\left. \left( {\frac{{{\rm d}{}}}{{{\rm d}{\,}}{r^{2}\psi }}}\right)
^{n+2}r^{2n+4}\rho (r)\,\right| _{r^{2}\psi \rightarrow h^{2}/2}\,,
\label{e:2.12}
\end{equation}
where specific energy and angular momentum are given by 
\begin{equation}
\varepsilon ={\textstyle{\frac{1}{2}}}(v_{r}^{2}+v_{t}^{2})-\psi \quad {\rm 
and}\quad h=r\,v_{t}.  \label{e: 1.1}
\end{equation}

Neglecting the self gravity of the star cluster in the vicinity of the
central black hole we write 
\begin{equation}
\psi (r)=\frac{GM}{r}\,,\quad {\rm and}\quad \rho (r)=\rho _{0}{\left( \frac{
r}{r_{0}}\right) ^{-5/2}}\,,  \label{e:powerlaw}
\end{equation}
where $M$ is the black hole mass and $G$ the Gravitational constant. The
slope of the cluster density profile is chosen to provide a compromise
between the observed m(K)$\leq $15 number counts (see below) and the
observed distribution of the innermost SgrA* cluster stars.

The resulting distribution from {Eqs.~(\ref{e:2.12})-(\ref{e:powerlaw})}
reads 
\begin{eqnarray}
f(\varepsilon, h) & \propto & (-\varepsilon)^{n+{\frac{1}{2}}} h^{2n-1} ={
\frac{3\,\sqrt{-\varepsilon}}{64\,h\,{{\pi }^3}}},{\frac{35\,{{\left(
-\varepsilon \right) }^{{\frac{3}{2}}}}\,h}{128\,{{\pi }^3}}}, {\frac{231\,{{
\left( -\varepsilon \right) }^{{\frac{5}{2}}}}\,{h^3}}{256\,{{\pi }^3}}}, {
\frac{1287\,{{\left( -\varepsilon \right) }^{{\frac{7}{2}}}}\,{h^5}}{512\,{
{\pi }^3}}}, {\frac{46189\,{{\left( -\varepsilon \right) }^{{\frac{9}{2}}}}
\,{h^7}}{7168\,{{\pi }^3}}}, {\frac{96577\,{{\left( -\varepsilon \right) }
^{{\frac{11}{2}}}}\,{h^9}}{6144\,{{\pi }^3}}}  \label{e:df} \\
& \propto & (-\varepsilon) [h/h_c(\varepsilon)]^{2n-1}
\end{eqnarray}
for $n=0,1,\cdots 5$. Here $h_c(\varepsilon)$ denotes the angular momentum
of the circular orbit at energy $\varepsilon$. The units are such that the
total mass and scale-radius, $r_0$, of the star cluster are unity and $GM=1$
. For this simple scale-free cluster in a Keplerian potential the
distribution functions are also derived in Sections 2.2 and 3.1 of Gerhard
(1991).

The velocity dispersion corresponding to {Eq.~(\ref{e:2.12})} is 
\begin{equation}
{\frac{\partial }{\partial r^{2}\psi }}\left( \sigma _{r}^{2}\rho
r^{2n+6}\right) =r^{2n+4}\rho ,  \label{e:betabeta}
\end{equation}
and together with the Jeans equation 
\begin{equation}
\frac{d}{{{\rm d}{r}}}\left( \rho \sigma _{r}^{2}\right) +\frac{2\beta }{r}
\rho \sigma _{r}^{2}=-\frac{\rho \,GM}{r^{2}}\,,  \label{e:jeansaniso}
\end{equation}
this implies 
\begin{equation}
\beta _{\star }=1/2-n\,,  \label{e:betastar}
\end{equation}
where the ${}_{\star }$ subscript refers to the fact that this is the {\sl 
intrinsic} anisotropy of the model. These simple models are therefore
scale-free and have constant anisotropy parameters $\beta _{\star }$, which
in the following will be chosen to match the range of values found for the
Galactic centre data set.{\vskip0.21cm}

\subsection{Monte Carlo star clusters}

\begin{figure*}
\unitlength=1cm 
\begin{picture}(12,8)
\put(-6,0){\centerline{\psfig{file=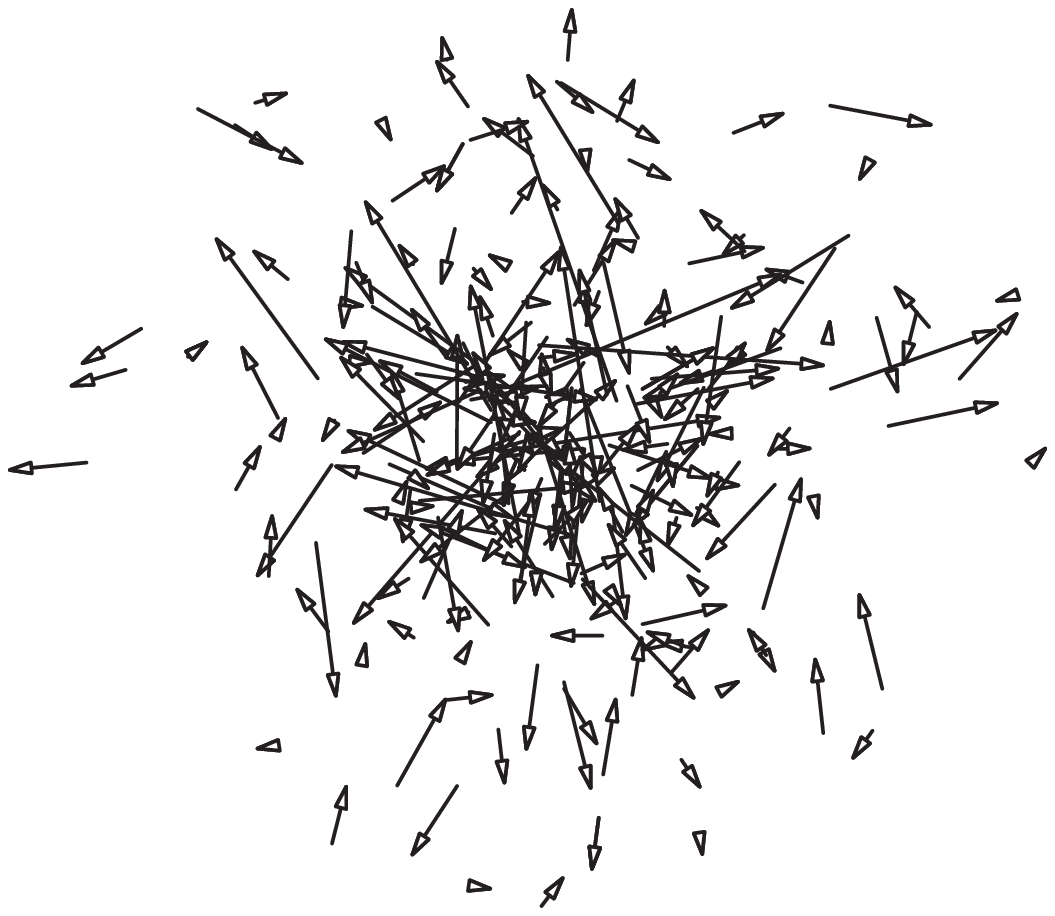,width=8cm}}}
\put(2,0){\centerline{\psfig{file=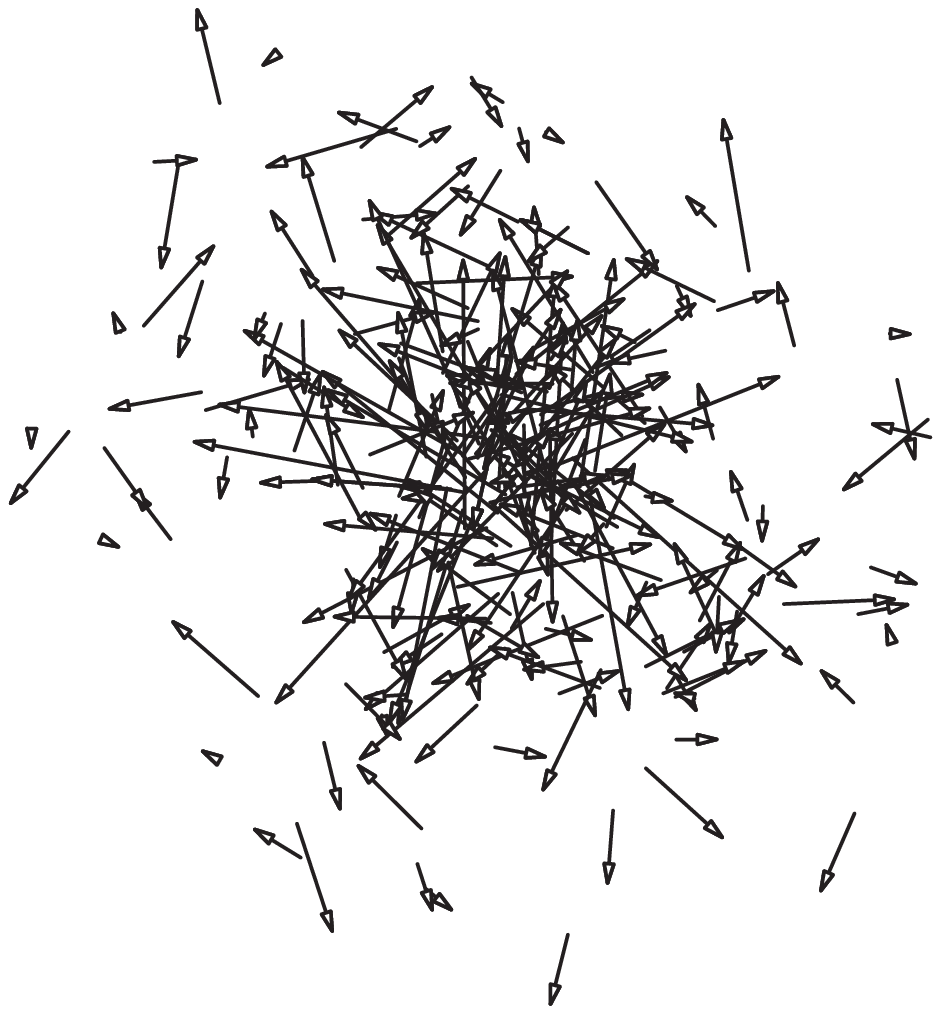,width=7cm}}}
\end{picture}
\caption{{\sl left panel: } Sky projection of 250 proper motions drawn from
a star cluster with $\protect\rho\propto r^{-2.5}$ and $\protect\beta_\star
=1/2$ (radial anisotropy) and {\sl right panel: } $\protect\beta_\star =-1/2$ 
(tangential anisotropy) in the potential
of a point mass. The distinction between radially and tangentially
anisotropic clusters is clearly visible. Radially anisotropic clusters
are more easily recognizable on the basis of the proper motion
patterns than tangentially anisotropic clusters. Note that it may appear incorrectly
that most stars are moving outwards because the inner plunging orbits have
longer vectors and seem to be moving radially outwards as the arrows
overshoot the centre. }
\label{f:genzel6}
\end{figure*}

\begin{figure*}
\unitlength=1cm 
\begin{picture}(10,8)
\put(-7,4){\centerline{\psfig{file=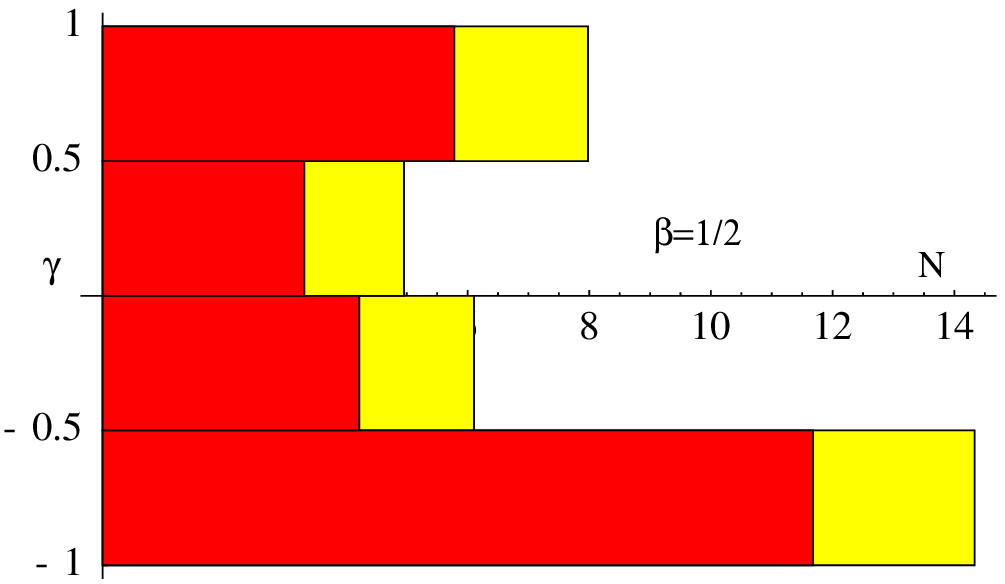,width=5.5cm}}}
\put(1,4){\centerline{\psfig{file=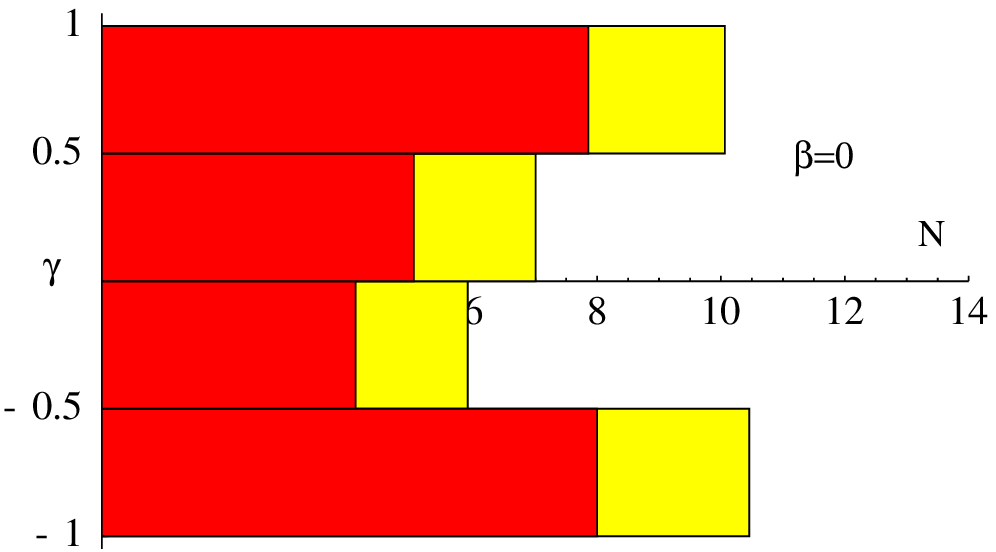,width=5.5cm}}}
\put(-7,0){\centerline{\psfig{file=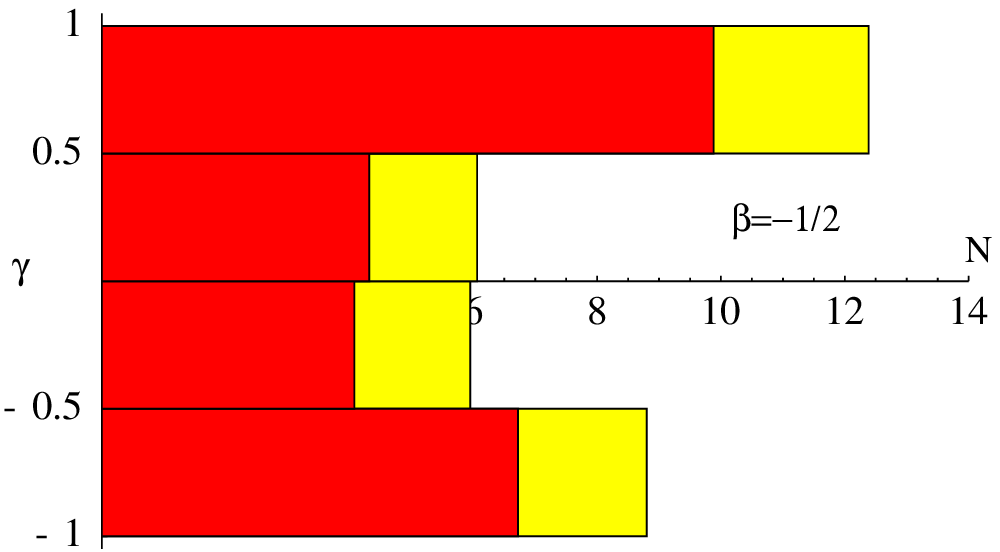,width=5.5cm}}}
\put(1,0){\centerline{\psfig{file=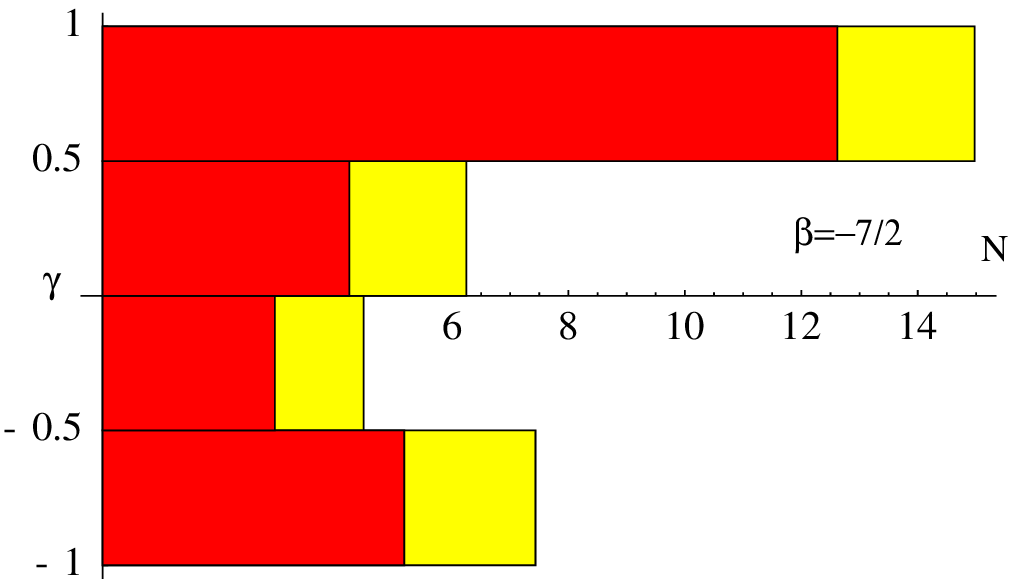,width=5.5cm}}}
\end{picture}
\caption{Histograms of $\protect\gamma_{TR} = (v_T^2
-v_R^2)/( v_T^2+v_R^2 )$ for a total number of 25 stars drawn from the
simulations described in Section \ref{s:mts}. Dark histograms show the mean
number of stars per bin averaged over 50 draws of 25 stars each out of a
5000 stars cluster. The added light histogram shows the mean relative errors
of the bin values. From top to bottom and left to right: $\protect\beta
_\star =0.5, 0, -0.5, -3.5$. Note that the isotropic model shows more stars
on projected tangential than on projected radial orbits. }
\label{f:genzel7}
\end{figure*}

\begin{figure*}
\unitlength=1cm 
\begin{picture}(7,6)
\put(-10,0){\centerline{\psfig{file=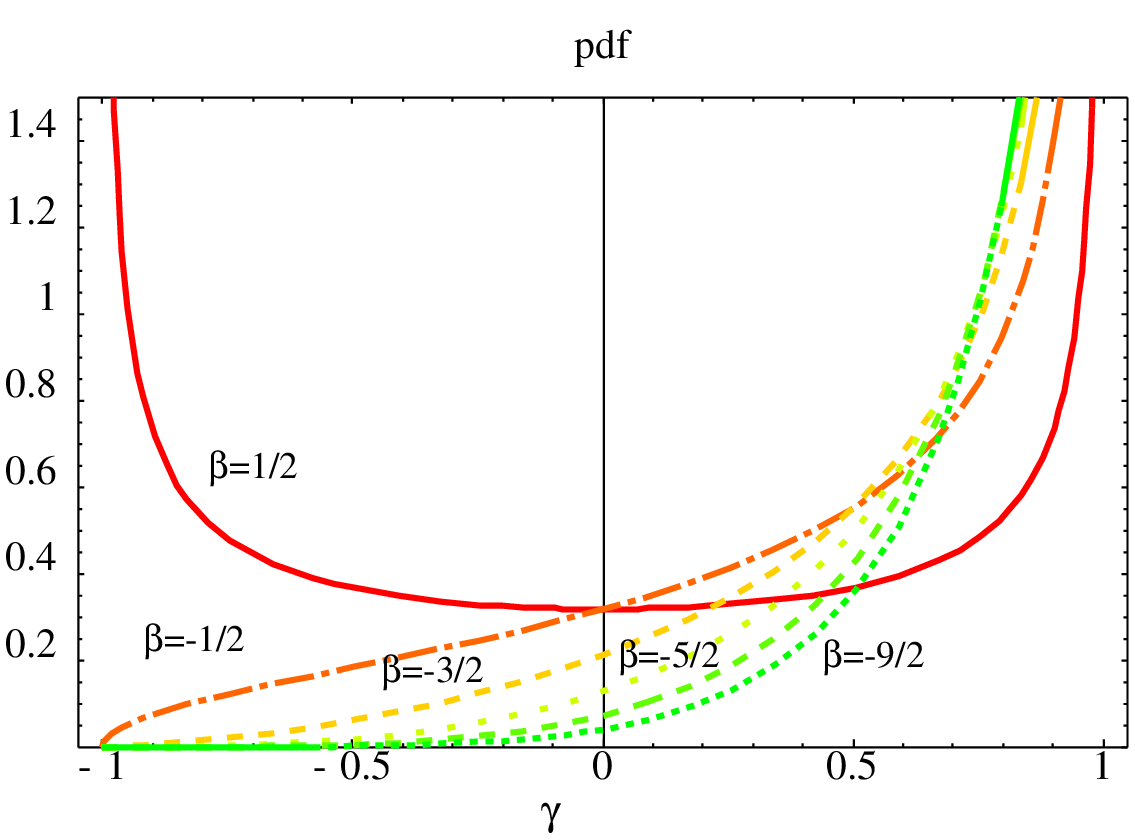,width=8cm}}}
\put(-1.5,0){\centerline{\psfig{file=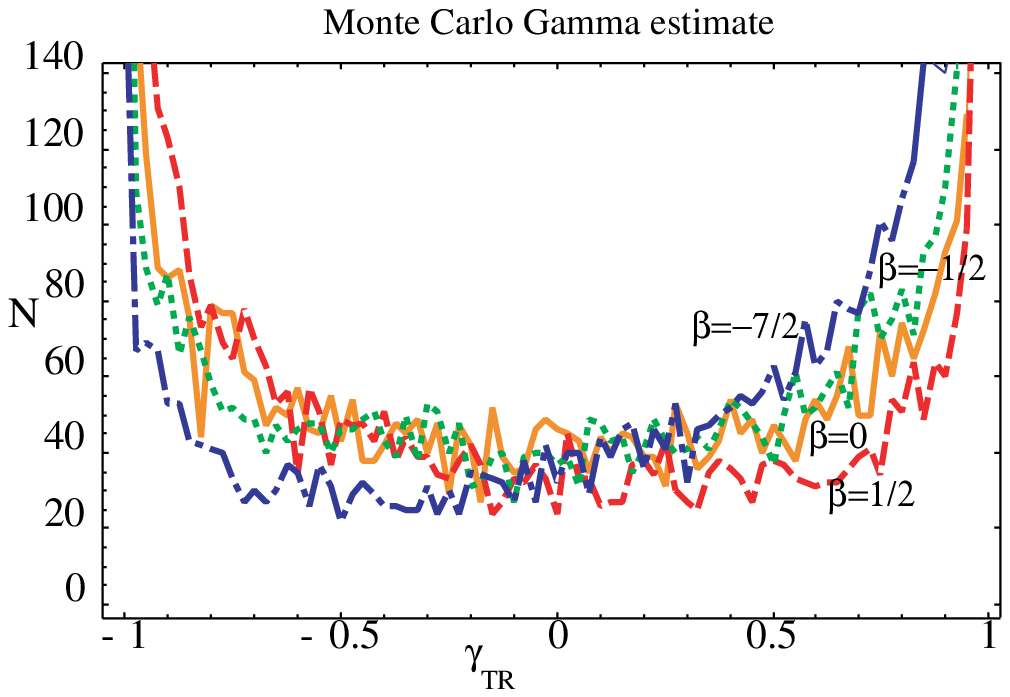,width=8.5cm}}}
\end{picture}
\caption{{\sl Left panel:} Theoretical $\protect\gamma_{tr}$ probability
distribution (pdf) for star clusters with constant anisotropy given by $
\protect\beta_\star=1/2, -1/2, -3/2, -5/2, -7/2, -9/2$, corresponding to
expectation values for $\protect\gamma_{tr}$ of $n/(n+1)=0,1/2,$ $2/3,3/4,4/5
$,and $5/6$ and variance of ${({1 + 2\,n})/ {{{( 1 + n ) }^2} \,/( 2 + n ) }}
={{1}/{2}}, {{1}/{4}},{{5}/{36}},{{7}/{80}}, {{3}/{50}} $ and ${{11}/{252}}
$. These distributions are very non Gaussian and peaked near $\pm1$. The $
\protect\beta_\star=1/2$ curve is symmetric (rather than that for the
isotropic model) because we have defined $\protect\gamma_{tr}$ in terms of
the total $v_t^2$ on the sphere rather than one half that. For these pdfs
the mean and standard deviation are well--defined for all values of $\protect
\beta_\star $. {\sl Right panel:} Projected $\protect\gamma_{TR}$ pdf for
2000 stars in clusters with $\protect\beta_\star=1/2, 0, -1/2, -7/2$
(dashed, full, dotted, and long dash-dotted lines). As for $\protect\gamma
_{tr}$, none of these curves is uniform. Note that all pdfs have significant
tails for $\protect\gamma_{TR}=-1$ because even purely tangential orbit may
project to projected radial orbits on the sky. Most of the difference
between the various models is in the relative number of stars in the radial
and tangential peaks near $\pm1$: The most tangentially anisotropic model
(long dash-dotted line) has the most stars for $\protect\gamma_{TR} \in
[0.5,1[$, whereas the radially anisotropic cluster (dashed line) is
overabundant in near radial proper motions. The curve for the isotropic
cluster is nearly symmetric. }
\label{f:genzel8}
\end{figure*}

\begin{figure*}
\unitlength=1cm 
\begin{picture}(20,16)
\centerline{\psfig{file=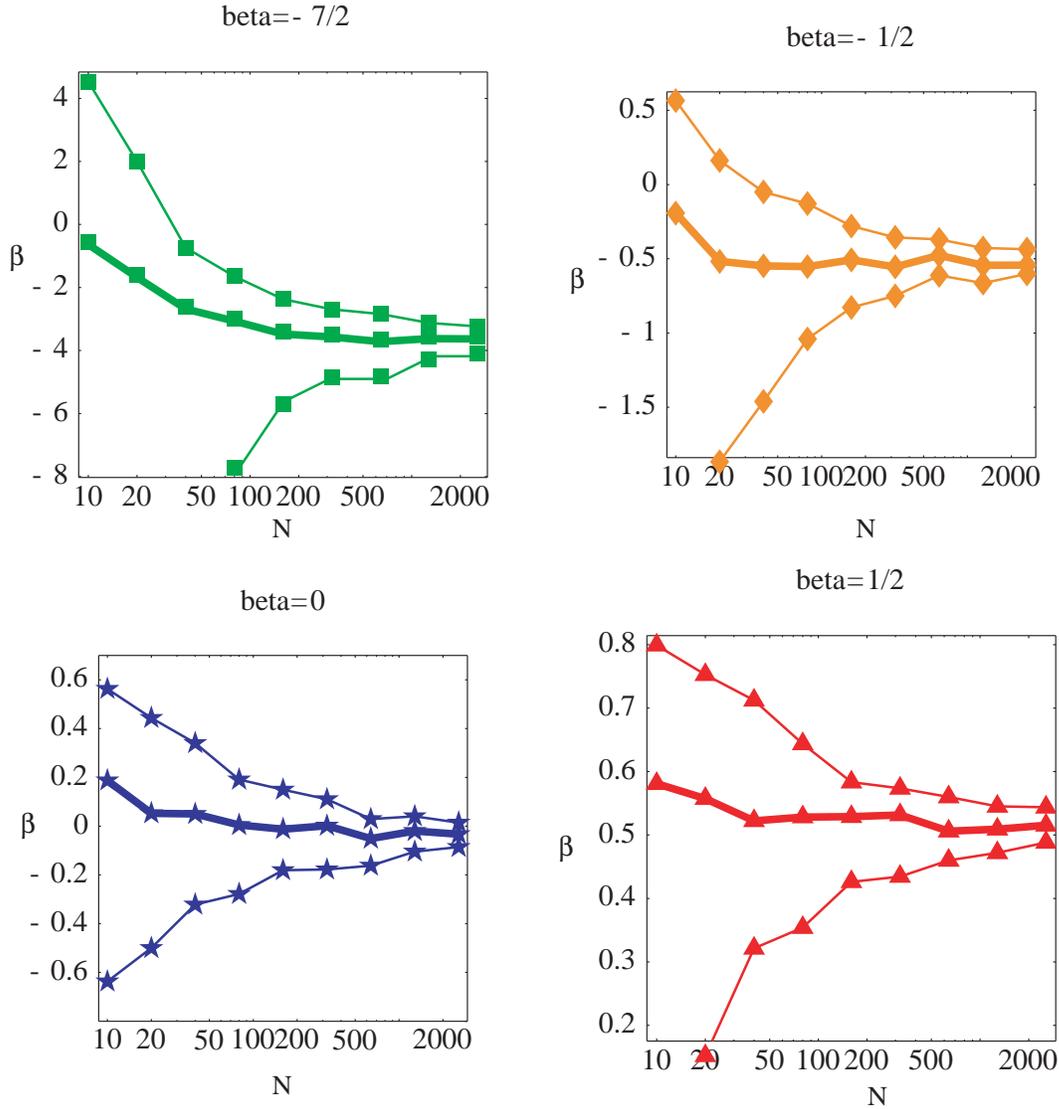,width=14cm}}
\end{picture}
\caption{Median (thick line) and first and third quartiles (thin lines) for
the distribution of $\langle \beta \rangle$ values determined by {Eq.~(\ref
{e:(6)})}, from simulated proper motion samples, as a function of sample
size (per bin) and for several values of true $\protect\beta_\star$ for the
underlying star cluster. The asymmetric width of the confidence bands
reflects the asymmetric nature of the anisotropy parameter $\protect\beta$.
For small samples {Eq.~(\ref{e:(6)})} allows unphysical derived values for $
\protect\beta$. Note the drift of the median as a function of $N$ which also
reflects the fact that the pdf of $\hat \protect\beta $ has inherited that
of $\protect\beta$, with both skewed towards negative values. See {Eq.~( \ref
{e:pdfbeta})} and {Fig.~(\ref{f:genzel10})}. The rather slow convergence of
the estimator as a function of the number of stars for models with more
negative $\protect\beta$ is striking; this follows indirectly from {Eq.~(\ref
{e:pdfbeta})} which has no well-defined moments.}
\label{f:genzel9}
\end{figure*}

\begin{figure*}
\unitlength=1cm 
\begin{picture}(20,8)
\put(-4.5,0){\centerline{\psfig{file=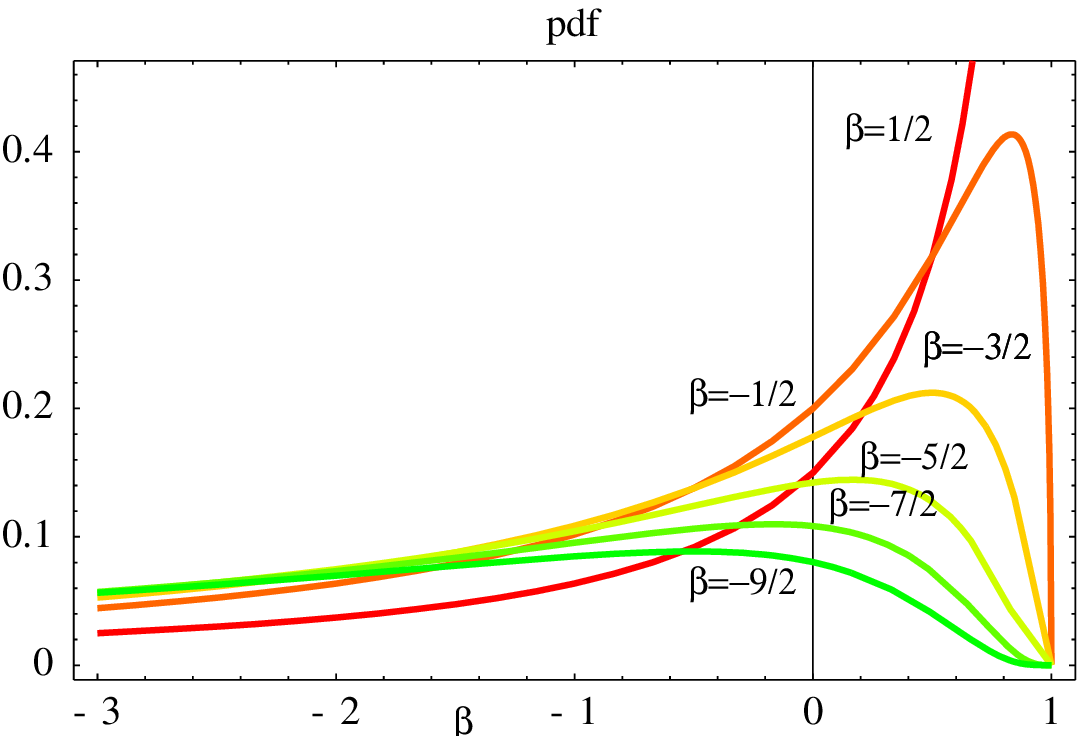,width=9cm}}}
\put(4.5,0){\centerline{\psfig{file=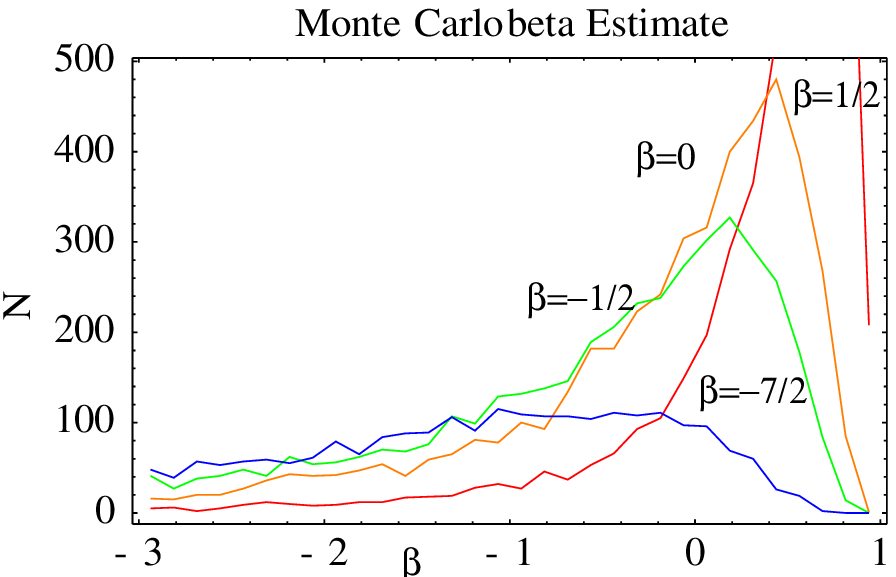,width=9cm}}}
\end{picture}
\caption{{\sl Left Panel:} Theoretical $\protect\beta$ probability
distribution (pdf) for $\protect\beta_\star =1/2, -1/2 \cdots-9/2$ as
labelled. These distributions are very skewed, and their means and standard
deviations are ill-defined. {\sl Right Panel:} Monte Carlo estimate of the
probability distribution of $\hat \protect\beta $ for models corresponding
to $\protect\beta_\star =1/2,0, -1/2$ and $-7/2$. The distributions result
from applying {Eq.~(\ref{e:(6)})} to the values of $\langle v_R^2 \rangle$
and $\langle v_T^2 \rangle$ obtained for many random realizations of 20
stars from the respective star cluster models. }
\label{f:genzel10}
\end{figure*}

We generate $N$ stars sampled regularly in radius with a cumulative mass
profile corresponding to {Eq.~(\ref{e:powerlaw})}. These are also required
to obey {Eq.~(\ref{e:df})}, {\it i.e.} the number of stars at radius $r$
within ${{\rm d}{r}}$ with radial velocity $v_{r}$ within ${{\rm d}{v}}_{r}$
and tangential velocity $v_{t}$ within ${{\rm d}{v}}_{t}$ is given by 
\begin{equation}
{{\rm d}{N}}=8\pi ^{2}\,r^{2}v_{t}\,f({\textstyle{\frac{1}{2}}}
(v_{r}^{2}+v_{t}^{2})-\psi (r),rv_{t}){{\rm d}{r}}{{\rm d}{v}}_{t}{{\rm d}{v}
}_{r}\,.  \label{e:dN}
\end{equation}
Given a triplet $(r,v_{r},v_{t})$, we generate a random position vector $
{\bf {r}}=r\{\cos (\theta )\cos (\phi ),\cos (\theta )\sin (\phi ),\sin
(\theta )\}$ where $z$ is along the line-of-sight as before, $\phi $ is a
random number uniform in $[0,2\pi \lbrack $ and $\sin (\theta )$ is a random
number uniform in $[-1,1[$. We also construct $v_{\theta }=v_{t}\cos (\chi )$
and $v_{\phi }=v_{t}\sin (\chi )$ where $\chi $ is a random number uniform
in $[0,2\pi \lbrack $. The velocity vector then reads ${\bf {v}}=v_{r}{\bf {
e_{r}}}+v_{\theta }{\bf {e_{\theta }}}+v_{\phi }{\bf {e_{\phi }}}$. It is
then straightforward to project the components of ${\bf {r}}$ and ${\bf {v}}$
onto the plane of the sky.{\vskip0.21cm}

{Fig.~(\ref{f:genzel6})} displays sky projections of the proper motion vectors
of 250 stars drawn from $\beta _{\star }=1/2$ and $-1/2$ clusters,
respectively. The length of each arrow is proportional to the magnitude of
the projected proper motion. The figure shows that a radially anisotropic
cluster is readily recognized by the many stars with radial proper motions:
intrinsically radial orbits remain radial when projected onto the sky, and
the number of radial proper motions is a good indicator of the number of
radial orbits. By contrast, intrinsically tangential orbits may appear
tangential or radial in the sky plane, depending on the orientation of their
orbital planes. Correspondingly, the projection of the tangentially
anisotropic cluster in {Fig.~(\ref{f:genzel6})}, while showing fewer radial and
more tangential proper motion vectors, still contains a significant number
of the former. Therefore, tangentially anisotropic clusters are more
difficult to recognize and discriminate from each other in terms of their
apparent proper motion distributions. Once the model is sufficiently
tangential, the ratio of radial to tangential proper motions is largely
determined by the projection rather than by the intrinsic anisotropy.{
\vskip
0.21cm}

\subsection{Anisotropy estimators}

From the Monte Carlo sample of stars we can estimate the previously used
anisotropy indicators $\gamma _{TR}$, and $\langle \beta \rangle$ ({Eq.~(\ref{e:(6)})
} ). The histograms of the estimated $\gamma _{TR}$ in {Fig.~(\ref
{f:genzel7})} confirm the above discussion quantitatively: (i) They show
that radially anisotropic models are more easily recognized by their proper
motion anisotropy than tangentially anisotropic models. Nonetheless,
strongly tangentially anisotropic clusters are recognizable in terms of
their many stars with $\gamma _{TR}$ near +1. (ii) The distribution of $
\gamma _{TR}$ is slightly skewed towards positive values even for
near-isotropic clusters. (iii) The histograms are always bimodal, i.e., have
peaks near $\gamma _{TR}=\pm 1$. This is also recognizable in the data; cf.\
the bottom right panel of {Fig.~(\ref{f:genzel4})}.

These histograms are discrete realizations of the probability distribution
for $\gamma _{TR}$, and this in turn derives from the marginal probability
distribution for the intrinsic quantity $\gamma _{tr}={(v_{t}^{2}-v_{r}^{2})}
/{(v_{r}^{2}+v_{t}^{2})}$, i.e., the number of stars with $\gamma _{tr}$ in
a small interval ${\rm d}\gamma _{tr}$. Once the distribution function is
known this is straightforward to compute: 
\[
{\rm pdf}(\gamma _{tr}){\rm d}\gamma _{tr}\propto \left\{ \int f\left(
\varepsilon (v_{r}[v_{t},\gamma _{tr}],v_{t}),rv_{t}\right) \frac{\partial
v_{r}}{ \partial \gamma _{tr}}\,v_{t}\,{{\rm d}{v}}_{t}\right\} \,{\rm d}
\gamma _{tr}\,, 
\]
which yields (after normalization) 
\begin{equation}
{\rm pdf}(\gamma _{tr}){\rm d}\gamma _{tr}={\frac{n!(\sqrt{1+\gamma _{tr}}
)^{2n-1} }{\pi \,(2n-1)!!\sqrt{1-{\gamma _{tr}}}}}{\rm d}\gamma _{tr}\quad 
{\rm where} \quad (n)!!=n(n-2)(n-4)\cdots 1\,.  \label{e:pdfgamma}
\end{equation}
This pdf is illustrated in {Fig.~(\ref{f:genzel8})}; it is strongly
non-Gaussian and skewed for both $n>0$ and for an isotropic cluster. The
reason why the isotropic curve is not symmetric is because we have defined $
\gamma _{tr}$ in terms of the total $v_{t}^{2}$ on the sphere rather than
one half that. The main point of this diagram is the non-uniform and
sometimes bimodal shape of the distribution. The distribution of the {\sl 
observed} $\gamma _{TR}$ pdf after projection is also shown in in {Fig.~(\ref
{f:genzel8})} as a histogram for 5000 stars. Relative to the intrinsic pdf
the number of (projected) radial orbits has been boosted, as discussed
above; the distribution is now always bimodal. Thus in diagrams like {Fig.~(
\ref{f:genzel5})} we should expect to always find an overabundance of stars
near $\gamma _{TR}=\pm 1$ compared to values near $\gamma _{TR}\simeq 0$,
with the ratio $N(\gamma _{TR} 
\mathrel{\spose{\lower
3pt\hbox{$\mathchar"218$}} \raise
2.0pt\hbox{$\mathchar"13C$}}1)/N(\gamma _{TR} 
\mathrel{\spose{\lower
3pt\hbox{$\mathchar"218$}} \raise 2.0pt\hbox{$\mathchar"13E$}}-1)$
containing the information about anisotropy.

{Fig.~(\ref{f:genzel9})} shows the median and first and third quartiles for
the distribution of $\langle \beta \rangle$ values determined by {Eq.~(\ref{e:(6)})}
from simulated proper motion samples, as a function of sample size and for
several values of true $\beta _{\star }$ for the underlying star cluster.
These confidence bands are especially wide for negative values of $\beta
_{\star }$ because it is a very asymmetric indicator of anisotropy. Indeed
the marginal propability distribution for the $\beta $ values as determined
from {\sl individual} stellar velocities is given by (following the
derivation of {Eq.~(\ref{e:pdfgamma})}) 
\begin{equation}
pdf(\beta ){{\rm d}{\beta }}={\frac{\sqrt{2}\,2^{2n}\,n!(\sqrt{1-\beta }
)^{2n-1 }}{\pi \,(2n-1)!!{(3-2\beta )^{n+1}}}}{{\rm d}{\beta }}\,.
\label{e:pdfbeta}
\end{equation}

This pdf is illustrated on the left panel of {Fig.~(\ref{f:genzel10})}. {
Eq.~( \ref{e:pdfbeta})} does not have any moments (i.e. the pdf does not
fall off fast enough as a function of $\beta $ to allow for, say the mean
and the variance to be computed). This implies that any estimator for its
central value will be unreliable. The pdf for $\langle \beta \rangle$ estimated via
Monte Carlo simulations, has inherited these asymmetries; see the right
panel of {Fig.~( \ref{f:genzel10})}. Because of the observed skewness we
expect the mean and the median to overestimate the anisotropy, especially
for more negative $\beta _{\star }$ models, as was indeed seen in {Fig.~(\ref
{f:genzel9})}.{\vskip0.21cm}

We are now in a position to discuss the inferred anisotropies of the
Galactic Centre star cluster ({Fig.~(\ref{f:genzel4})} and Table~2) in more
detail. Comparing with the distributions in {Fig.~(\ref{f:genzel8})}, the
evidence for radial anisotropy in the central 0.''8 rests on the absence of
stars with $\gamma_{TR}\simeq 1$, and the case for the tangential anisotropy
of the HeI stars on the absence of stars with $\gamma_{TR}\simeq -1$. Based
on the Monte Carlo models the  evidence for anisotropy of the orbits is
fairly solid. With larger proper motion samples, it may be best to compare
with these distributions directly to estimate the anisotropy. The values for 
$\beta$ estimated from {Eq.~(\ref{e:(6)})}, on the other hand, are quite
uncertain. With this estimator being a quotient of observable dispersions,
its distributions are very broad ({Fig.~(\ref{f:genzel10})}). The Monte Carlo
simulations indicate that a sample of 500 stars with $v/\sigma (v)$ greater
than 3 will be required in order to determine even $\beta _{\star }=1/2$ to
an accuracy of $\pm 0.2$.

In summary, the number of observed stars and the quality of the derived
velocities is already sufficient to state with some certainty that
anisotropies in the orbits of (early type) stars are indeed present. To be
consistent with the observed distribution of $\gamma _{TR},$ the model
clusters (assuming sphericity and cylindrical
symmetry of the velocity ellipsoid) 
require fairly strong radial anisotropy at small radii, and
tangential anisotropy for larger radii. However, the data are not yet suited
to place accurate quantitative constraints on the anisotropy parameter $
\beta $ and its radial dependence. {\vskip0.21cm}

\section{ Global Rotation of the Early Type Cluster}

\begin{figure}
\centering \includegraphics[width=0.95\columnwidth,clip=true]{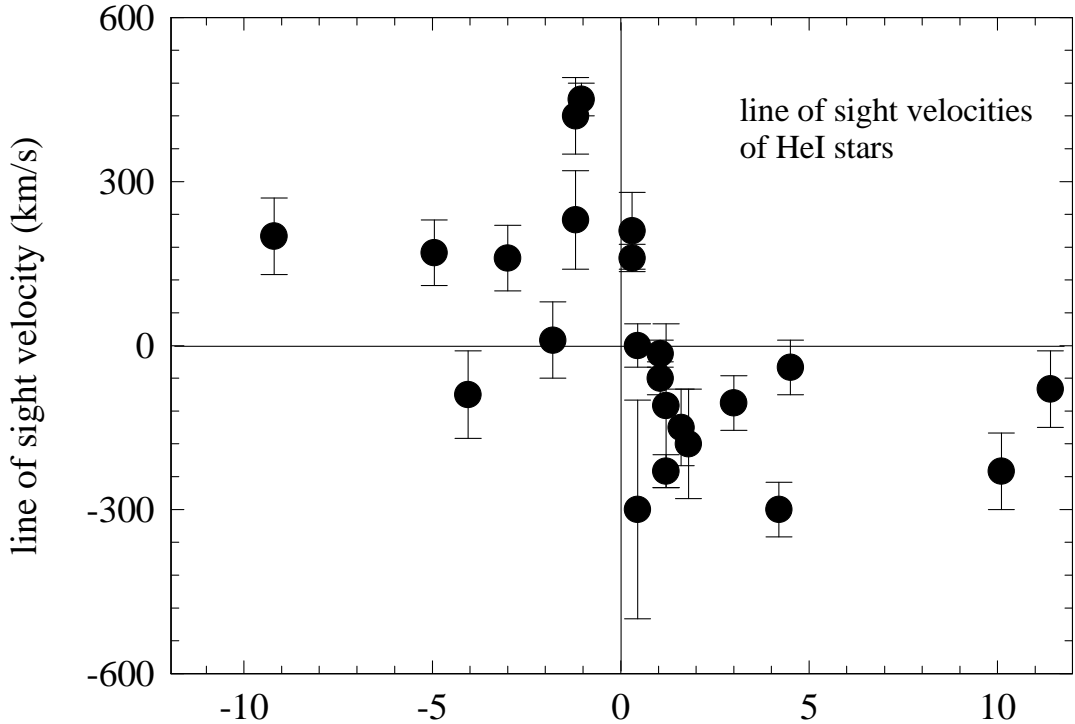}
\caption{ Line of sight velocities of all HeI stars in Table~1, as a
function of Dec-offset from SgrA*. Error bars a $\pm 1 \protect\sigma $.}
\label{f:genzel11}
\end{figure}

\begin{figure}
\vskip 0cm \centering \includegraphics[width=0.85\columnwidth]{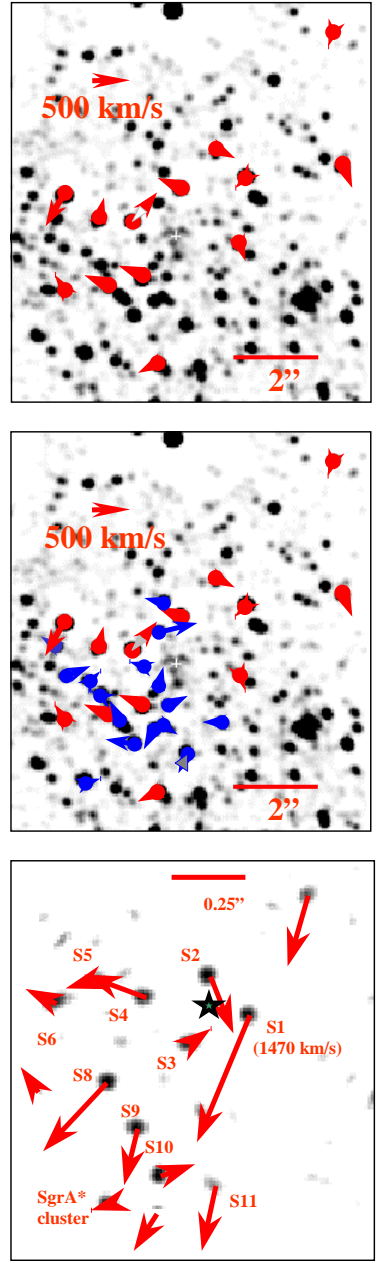}
\vskip -2.0cm
\caption{  Proper  motion  vectors on  K-band  images  of  the central  star
cluster. North is up and East is to the left. Top: proper  motion vectors of  all HeI  stars  with three
velocities, overlayed on a grey scale  SHARP map at $0.15``$ resolution. The
length and direction of  each arrow denotes  the magnitude and  direction of
the proper  motion for  each star. For   comparison a  500 km/s motion  (for
$R_{{ \sun}}=8.0 {{\rm kpc}}$) is shown in the upper  left. To indicate the
uncertainties in direction for  the best stars, a  shaded cone is  shown for
IRS 16C. The white cross denotes the location of SgrA*.  The middle panel shows in
addition  the vectors   for  all stars  in  the   IRS 16 complex  that  have
K-magnitude  $\leq 13$. This  criterion  likely selects primarily early type
stars. Proper  motion vectors within $\approx {}0.7``$  of SgrA* (the `SgrA*
cluster`), overlayed on  the  0.05`` resolution  K-band  map of Ghez et  al.
(1998) are given in the bottom panel. The asterisk marks the position of
SgrA*. The $1 \protect\sigma $ positional uncertainty of the radio source on
the  infrared map  is    between $\pm   $10  and $\pm    $30 milli-arseconds
(1$\protect  \sigma $). The identifications  of the sources (S1-S11,
Table~1) are  given.  The length of  the proper motion  of S1  corresponds to 1470
km/s for $R_{{\sun} }=8.0{{{\rm kpc}}}$. }
\label{f:genzel12}
\end{figure}

As a group, the early type stars (= the starburst component) exhibit a well-defined
overall angular momentum. The line-of-sight velocities of the 29
emission line stars follow a rotation pattern: blue-shifted radial
velocities north, and red-shifted velocities south of the dynamic centre ({
Fig.~(\ref{f:genzel11})}). The apparent rotation axis is approximately
east-west, within $\pm $20$^{{\circ}}$. The early type stars thus are in a
counter-rotation with respect to general Galactic rotation, the latter
showing blue-shifted material south and red-shifted material north of the
Galactic centre. The rotation is fast (average $\approx ${}150 km/s) and is
consistent with a Keplerian boundary for a 2 to 3 million solar mass central
mass ({Fig.~(\ref{f:genzel11})}). Our results confirm and improve the earlier
conclusions of Genzel et al. (1996). Note that the late type stars also show an
overall rotation, but that is slow (few tens of km/s) and consistent with
Galactic rotation (McGinn et al. 1989, Sellgren et al. 1990, Haller et al.
1996, Genzel et al. 1996).{\vskip0.21cm}

Eckart and Genzel (1996) have argued that the HeI stars also show a
coherent pattern in their proper motions. Such a pattern is now confidently
detected in the data ({Fig.~(\ref{f:genzel12})}). It is the origin for
much of the tangential anisotropy discussed above. 11 of the 13 proper motion
vectors for the emission line stars display a clockwise pattern, with only
IRS16 NE and IRS 16NW moving counter-clockwise. A number of authors have
argued that most of the members of the IRS16 complex (located between SgrA*
and 4'' east of it, and between 3.5'' south of SgrA* and 1.5'' north of it)
belong to the early type cluster, with the HeI stars just being a sub-sample
of the brightest emission line objects. This assertion is confirmed as well.
Most of the brighter stars in the IRS16 complex (m$_{{K}}\leq $13) show a
clock-wise streaming pattern ({Fig.~(\ref{f:genzel12})} middle panel). In {
Fig.~(\ref{f:genzel12})} (bottom panel) we overlay the proper motions vectors
of Table~1 on the 0.05'' resolution K-band Ghez et al. (1998) image of the
SgrA* cluster. The preference of stars to be on radial/highly elliptical
orbits that was discussed in the last section can be checked here from a
graphical representation. {Fig.~(\ref{f:genzel12})} also suggests that the
majority of stars in the SgrA* cluster have a similar projection of angular
momentum along the line-of-sight, L$_{{z}}$, as the much brighter HeI
stars and the IRS 16 cluster members. Speckle spectrophotometry (Genzel et
al. 1997) and very recent high resolution VLT spectroscopy (Eckart et al.
1999) show that the brighter members of the SgrA* cluster lack 2.3-2.5$\mu $
m CO overtone absorption features and thus are clearly not late type stars.
They are probably early type stars. If they are on the main sequence they
would be of type B0 to B2. {\bf {\it The SgrA* cluster members thus are
probably part of the early type star cluster but are on plunging, radial or
very elliptical orbits}}. The only alternative explanation of the observed
radial anisotropies and net angular momentum is that the SgrA* cluster
stars are rotating as a group, like the HeI stars, but with a rotation
axis lying in or near the plane of the sky. While this explanation seems
relatively implausible, measurements of proper motion curvature and
radial velocities are required to make a decisive test.

However, the early type cluster cannot simply be modeled as an inclined, rotating
thin disk. The fit of the best Keplerian disk model (inclination 40$^{{o}}$
, v$_{{rot}}$=200 r$^{{-0.5}}$) to the HeI star velocities is poor.
There are no HeI stars seen at Galactocentric radii greater than \symbol{126}
12''. Because the HeI stars should be phase-mixed along their orbits (\S
2.2), a better description of their distribution, and perhaps the entire
early type cluster, probably is a dynamically hot and geometrically thick,
rotating torus at radii from 1'' to 10'' (0.039 to 0.39 pc).

Most of the stars in the torus will have a fairly large angular momentum L
and approximately the same sign of L$_{{z. }}$ In the distribution of
different L's there is a small fraction of stars, however, with much smaller
L and still the same sign(L$_{{z}}$). This sub-population is necessarily
small and may thus not contain very massive stars. The low-L stars are able
to pass much closer to SgrA* than the majority of the early type star
cluster. In our interpretation, it is these stars on largely radial,
plunging orbits that make up the SgrA* cluster. As the bright, more massive
stars are on average at larger distances from SgrA*, it is possible to
detect fairly easily this central sub-sample of fainter,
fast moving stars. One would expect to find the same types of stars also at
larger radii from SgrA*. However, the present proper motion data sets are
biased against such fainter stars because of the presence
of the brighter early type stars (especially the IRS 16 cluster) and of late
type stars at yet larger true radii. 

In summary of this section, we conclude that the majority of the HeI
emission line stars and the bright (early type) stars in the IRS 16 cluster
show a coherent clockwise and counter-Galactic rotation. Their circular
(tangential) velocities dominate over their radial velocities. The young
stars are arranged in a thick torus of mean radius $\approx${}0.2 pc. This
torus was presumably first formed $\approx${}7 to 9 million years ago when
one or several infalling, tidally disrupted clouds collided and were
highly compressed. This lead to an episode of active star formation in the
central parsec. From the presence of bright AGB stars in the same region
(Krabbe et al. 1995, Genzel et al. 1996, Blum, Sellgren and dePoy 1996) it
is likely that there were other such phases of active star formation in the
more distant past (a few 10$^{{2}}$ million years ago).

\section{\bf Projected mass estimator and Anisotropy.}

 \begin{figure}
\centering \includegraphics[width=0.75\columnwidth]{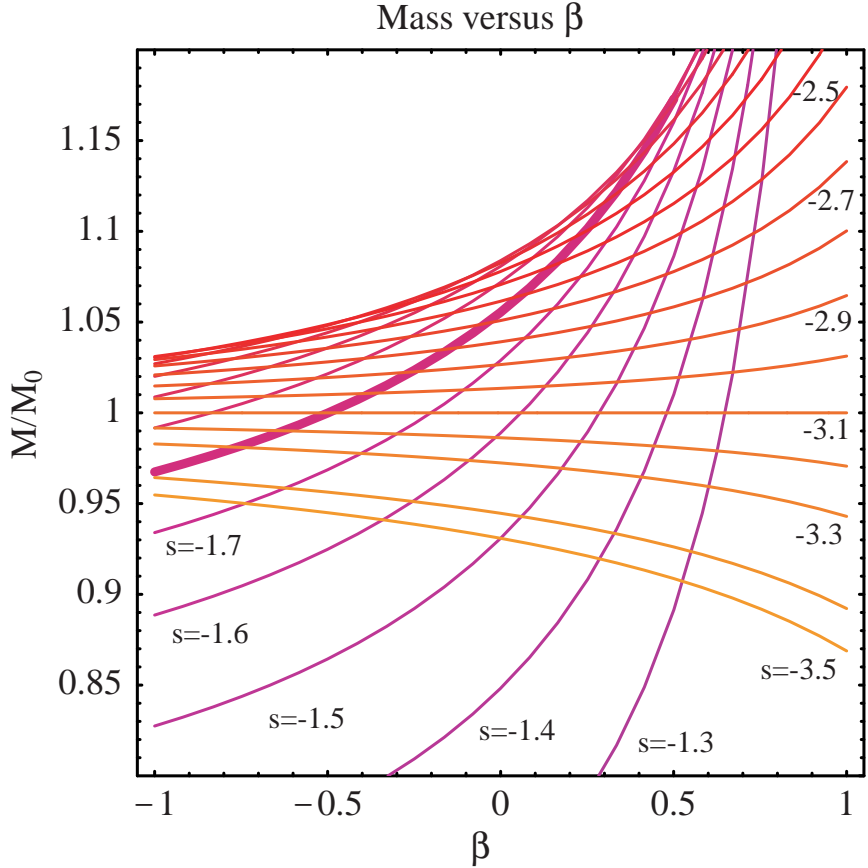}
\caption{ 
Leonard-Merritt mass  estimates (in units of the  true underlying mass) as a
function of $\beta$ for  different power law  slopes of clusters as labelled
(the  thicker curve corresponds  to $\rho \propto  r^{-1.8}$). Note that for
all power law clusters but that  corresponding to $\rho \propto r^{-3}$, the
mass estimate is  typically biased  and  gives an overestimate or  underestimate 
of the
real  mass.  In  particular    note  that the  Bahcall   Tremaine   estimate
(corresponding to  $\beta =0$)  is also  offset for  all slopes but  $3$ and
$1.62$.  For instance  for  $\beta=0$   and  $\rho \propto  r^{-1.8}$,   the
relative mass discrepency reads $ M_{{LM}}/M_0 \approx  1.05$ but for $\beta
= -5$ it reaches $ \approx 0.89$ while a purely radial cluster would lead to
an overestimation of 60 \% as shown in Table~3.  In practice our estimate
of the mass of the  black hole should be rescaled  downward by about 5-10 \% to account
for this bias in the relevant radial range.  }
\label{f:correct}
\end{figure}

\begin{table}
{{\bf Table~3.} Correction factor ${M_{{\rm LM}}}/{M_0 }$ for LM mass estimator versus 
      $\beta$ (horizontally) 
      and $s$ (vertically). Note the steep rise near $\beta =1$ for 
      shallower density profiles. The function (${M_{{\rm 
      LM}}}/{M_0 })(\beta,s)$ is illustrated in \Fig{correct} } 
      \centering
      \begin{tabular}{|r|rrrrrrrrrr|}   \hline
	   $ s\setminus \beta $ & -5 & -4 & -3 & -2 & 
    -1 & 0 & 0.25 & 0.5 & 0.75 & 1 \\ \hline
   -1.2 & 0.43 & 0.43 & 0.44 & 0.46 & 0.49 & 0.57 & 0.62 & 0.71 & 0.93 & 2.3 \\ 
   -1.4 & 0.67 & 0.67 & 0.69 & 0.71 & 0.75 & 0.85 & 0.91 & 1. & 1.2 & 2. \\ 
   -1.5 & 0.74 & 0.75 & 0.77 & 0.79 & 0.83 & 0.93 & 0.99 & 1.1 & 1.3 & 1.9 \\
   -1.6 & 0.81 & 0.81 & 0.83 & 0.85 & 0.89 & 0.99 & 1. & 1.1 & 1.3 & 1.8 \\ 
   -1.7 & 0.85 & 0.86 & 0.88 & 0.9 & 0.93 & 1. & 1.1 & 1.2 & 1.3 & 1.7 \\ 
   -1.8 & 0.89 & 0.9 & 0.91 & 0.93 & 0.97 & 1.1 & 1.1 & 1.2 & 1.3 & 1.6 \\ 
   -1.9 & 0.92 & 0.93 & 0.94 & 0.96 & 0.99 & 1.1 & 1.1 & 1.2 & 1.3 & 1.5 \\ 
   -2. & 0.94 &  0.95 & 0.96 & 0.98 & 1. & 1.1 & 1.1 & 1.2 & 1.3 & 1.4 \\ 
   -2.2 & 0.97 & 0.98 & 0.99 & 1. & 1. & 1.1 & 1.1 & 1.1 & 1.2 & 1.3 \\ 
   -2.4 & 0.99 & 1. & 1. & 1. & 1. & 1.1 & 1.1 & 1.1 & 1.2 & 1.2 \\ 
  -2.5 & 1. & 1. & 1. & 1. & 1. & 1.1 & 1.1 & 1.1 & 1.1 & 1.2 \\ 
   -2.6 & 1. & 1. & 1. & 1. & 1. & 1.1 & 1.1 & 1.1 & 1.1 & 1.1 \\ 
   -2.8 & 1. & 1. & 1. & 1. & 1. & 1. & 1. & 1. & 1. & 1.1 \\ 
   -3. & 1. & 1. & 1. & 1. & 1. & 1. & 1. & 1. & 1. & 1. \\ 
   -3.5 & 0.98 & 0.98 & 0.98 & 0.97 & 0.95 & 0.93 & 0.92 & 0.91 & 0.89 & 0.87 \\ 
   \hline
   \end{tabular}
     \label{t:tab1}
 \end{table}

Leonard    and Merritt (1989)  have   shown  that an anisotropy-independent,
projected  mass estimator can  be constructed  from  {\sl radially complete}
proper motion data.  Starting   from the Jeans  equation for   a spherically
symmetric, non-rotating system,
\begin{equation}
GM(r)=-r\sigma _{{r}}^{{2}}\{{{\rm d}{\log }}n(r)/{{\rm d}{\log }}r+{{\rm d}{
\log }}\sigma _{{r}}^{{2}}/{{\rm d}{\log }}r+2(1-\sigma _{{t}}^{{2}
}/\sigma _{{r}}^{{2}})\}\quad ,  \label{e:(9)}
\end{equation}
one can construct the spatially averaged, stellar tracer density ($n ( r )$)
weighted expectation value $G \langle M( r )\rangle $. This estimator,
henceforth referred to as the Leonard-Merritt (LM) estimator, is independent
of any assumptions about anisotropy, 
\begin{equation}
G\langle M(r)\rangle _{{LM}}=\langle r(2\sigma _{{r}}^{{2}}+2\sigma _{{
t}}^{{2}})\rangle =(16/3\pi )\{2\langle R\sigma _{{R}}^{{2}}\rangle
+\langle R\sigma _{{T}}^{{2}}\rangle )\quad .  \label{e:(10)}
\end{equation}

Table~2 lists the LM-estimator obtained from 95 proper motion stars within $
5^{\prime\prime}$ of SgrA*. The estimated mass is $2.9\pm 0.4\times 10^{{6}} M_{\sun 
}$. It  is quite insensitive to  the weighting scheme of   the data. We also
list LM-estimates for  different projected  annuli  (although this is
formally not appropriate, see below).

For  comparison, the Bahcall-Tremaine (1981,  BT) estimator for an isotropic
cluster around a point mass gives a mass of $3.1\pm 0.3\times 10^{{6}} M_{
\sun }$ for the 2x95 proper motions within $R=5^{\prime\prime}$ (Table~2).
For  purely radial orbits the  mass  would be  twice   as large. Note
however that formally the BT estimator is defined for radial
velocities only and as such the application to proper motions is
inappropriate.  The  virial
theorem mass  estimate (VT) of the same  data gives  $2.5-2.6\times 10^{{6}}
M_{\sun }$ (Table~2,  see Bahcall and Tremaine  1981, or Genzel et al. 1996
for a discussion). The BT estimator requires  prior knowledge of the
orbit structure. In the region outside 3'' from  SgrA* where from our proper
motion analysis  the orbit   structure   is approximately   isotropic,   the
agreement between all three  estimators is fair   (at somewhat more than  $
1\sigma$).

  \subsection{Correction for the  Leonard-Merritt mass estimate} 

Unfortunately, the Leonard-Merritt mass estimate assumes that the cluster is
of finite mass and  that  we have access to   the full radial extent of  the
cluster. Here the density profile behaves  roughly as a  power law over the
finite range of radii for which data is available. For  such a mass model in
a Keplerian potential the  implementation of Leonard-Merritt  mass estimate
on concentric  rings yields a  biased (systematically offset) measure of the
mass.  Indeed the derivation of   this estimator involves an integration  by
part of integrals of the form
\begin{equation}
\int \limits_{0}^{\infty}  r^{4} \frac{{\rm d }\rho \sigma^{2}(r)}{{\rm d}r} {\rm 
d } r =- 4 \int \limits_{0}^{\infty} \rho(r) r^{3}
\sigma^{2}(r) {\rm d } r +\left[ \rho(r) r^{4} \sigma^{2}(r) 
\right]_{0}^{\infty}
\EQN{part} \end{equation}
For finite mass systems the second term of the right  hand side of \Eq{part}
vanishes. In the context of the Galactic centre we still compute
\begin{equation}
{\int_{r_{1}}^{r_{2}}   \rho(r)   r^{3}    \sigma^{2}(r) {\rm      d  }    r
}/{\int_{r_{1}}^{r_{2}} \rho(r) r^{2} \sigma^{2}(r) {\rm d } r }
\EQN{crap} \end{equation} 
over a finite radial range $r_{1}, r_{2}$ and even though both numerator and
denominator diverge as $r_{1}\rightarrow 0$ and $r_{1}\rightarrow
\infty$, the ratio is well defined and equals the value corresponding 
to finite $r_{1}$ and $r_{2}$. On the other hand
\begin{equation}
\left[ \rho(r) r^{4} \sigma^{2}(r) 
\right]_{r_{1}}^{r_{2}} / {\int_{r_{1}}^{r_{2}} \rho(r) r^{2} \sigma^{2}(r) {\rm d } r }
\EQN{crap} \end{equation}
is also finite but  non zero except for  $s=-3$.  As a consequence the ratio
$M_{{\rm  LM}}/M_0$ will  typically be  a function  of  $\beta$ and $s$, the
slope of the local power  law corresponding to  the range for which data  is
available. A straightfoward calculation yields
\begin{equation}
\frac{M_{{\rm LM}}}{M_0 }={\frac{{2^{4 + s}}\,{\Gamma}({{-s}/{2}})\,
     \left( \left(  3   -  s  \right)  \,\left(  1  -  s   -  \beta  \right)
     \,{\Gamma}(-s) - {\Gamma}(2-  s) \right) }{3\,{\sqrt{\pi }}\,\left( 1 -
     s  - 2\,\beta   \right) \,{\Gamma}({{ -s/2-1/2}{}})\,  {\Gamma}({{1/2 -
     s/2}{}})\,{\Gamma}({{3/2- s/2}{}})}}
\EQN{corect} 
\end{equation}
Relevant relative mass estimates   versus  $\beta$ for different power   law
index are shown in \Fig{correct}.  In practice \Eq{corect} is
used to correct for the offset in the measured $M_{{\rm LM}}$. Table~3 gives
a few values relevant for the Galactic centre.

Note that ${M_{{\rm LM}}}={M_0 }$  for $\beta =0$ when $s$   is $-3$ or  the
 root of 
\begin{equation} 16\,\Mfunction{\Gamma}({{s}/{2}})\,\left( \left( 1
 - s   \right) \,\left(   3    -    s \right)  \,\Mfunction{\Gamma}(s)     -
 \Mfunction{\Gamma}(2 -    s)   \right)  =   3\,{2^{1  -     s}}\,{\sqrt{\pi
 }}\,\Mfunction{\Gamma}({{   -s/2-1/2}{}})\,   {{\Mfunction{\Gamma}({{3/2  -
 s/2}{}})}^2} \EQN{crap} 
\end{equation} 
which yields $s \approx -1.62$. More generally there is  a non trivial curve
 (i.e. which differs from  ${s=-3}$) corresponding to $M_{{\rm  LM}}/M_0 =1$
 in the $(\beta,s)$ plane.
   
We conclude that the LM-estimator is not independent of  $\beta$ or $s$ when
applied to truncated data set  even though each shell  yields the same  mass
estimate for a Keplerian potential.  We do need to estimate $\beta$ and $s$
independently to correct  for the offset.  Since $\beta$  varies with radius
for the Galactic centre the correction will affect the mass profile.

\begin{figure*}
\unitlength=1cm 
\begin{picture}(20,10)
\put(-4,0)\centerline{\psfig{file=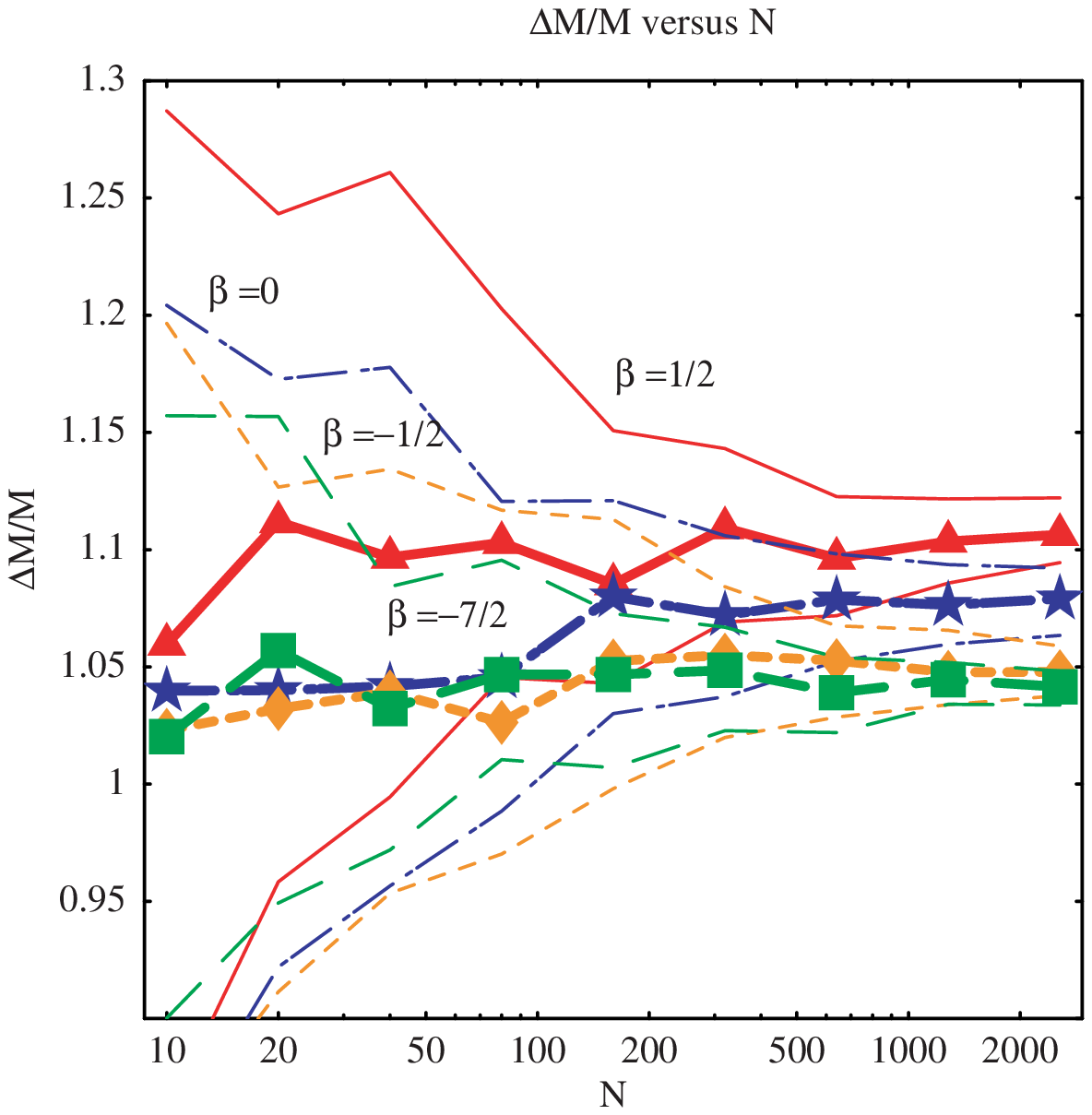,width=8cm}}
\put(4,0)\centerline{\psfig{file=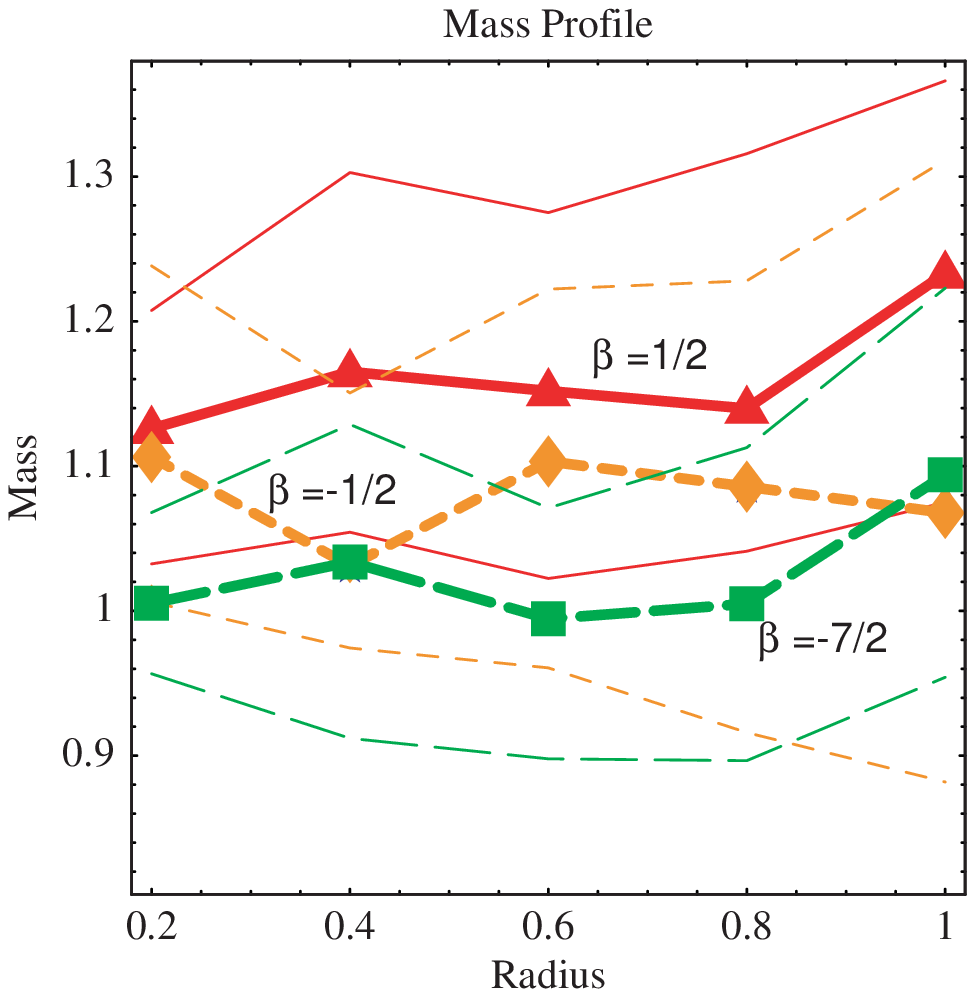,width=8cm}}
\end{picture}
\caption{Left: Median and quartiles for the distribution of values for the
 mass estimator $M_{LM}$ ({Eq.~(\ref{e:(6)})})  from simulated proper motion
samples, as a function of sample size  and for several underlying anisotropy
values    $\protect\beta_\star=0.5,0,-0.5,-3.5$  of  the   simulated    star
cluster. The true central  mass is unity  and the offset  induced by the  LM
estimation is clearly visible   there and in accordance with the theoretical
prediction from \Tab{tab1}. Mass determinations reliable at   the
10\% level  require  samples  of at  least 40,   35,  25, and 15   stars for
$\protect\beta_\star=0.5,0,-0.5,-3.5$.    Right:  Central    mass  estimates
$M_{LM}$ derived in five annuli on the sky for  a total number of 800 stars.
anisotropy values  are  $\protect\beta_\star=   0.5,-0.5,$ and $-3.5$,    as
labelled         (plain,dash-dot,long         dash,dash)   corresponding  to
(triangle,star,diamond,square).}
\label{f:genzel13}
\end{figure*}

The mass estimators are derived by averaging over the entire star cluster,
while the observed stars in the Galactic centre are presumably part of a
more extended stellar system. To test for possible systematic effects, we
have therefore also carried out Monte Carlo simulations for the
LM-estimator. Again, as in Section 4.2, we have used a power law
distribution of tracer stars with $n\propto r^{-2.5}$ as for the
kinematically measured stars. The left panel of {Fig.~(\ref{f:genzel13})} shows
the median and quartiles of the distribution of $M_{LM}$ values derived 
 for many star cluster realizations, as a function of sample size
and for $\beta _{\star }=0.5,0,-0.5$, and $-3.5$. The true mass of the
central black hole that dominates the potential of these clusters is $M=1$.
The right panel of {Fig.~(\ref{f:genzel13})} shows, for $N=800$ stars, the
effect of estimating the central mass from five concentric annuli aranged
linearly as a function of radius. Note that the mass profile is indeed flat 
(within the statistical uncertainties) as expected for a Keplerian potential
and offset by the amount predicted by \Eq{corect}.

These simulations suggest that applying the  LM-estimator to a
central sub-volume of the actual star  cluster around the black hole gives
the correct  hole   mass if the distribution  of   orbits  is strongly
tangential, independent of power law slope
 (all     radial   shells  should    then   be     independently
sufficient).   For  isotropic   and   radially  anisotropic  orbit
distributions and power law slopes near -2
the LM-estimator gives somewhat biased (too high) values for the central
mass. The value of $M=2.9
\times 10^6 M_\odot$ derived for the central mass from  all stars inside 5''
(an approximately overall isotropic  sample) thus is likely systematically
high by about $5-10\%$.

\subsection{{\bf Estimate of the Sun-Galactic centre distance R$_{{\sun}}$}}

The expectation   values of the   first  moments of the  projected  velocity
dispersions are related to each other through their mutual dependence on the
intrinsic radial and tangential velocity dispersions. One can write
\begin{equation}
\langle R\sigma _{{z}}^{{2}}\rangle =(1/3)\langle R\sigma _{{R}}^{{2}
}\rangle +(2/3)\langle R\sigma _{{T}}^{{2}}\rangle \quad .
\label{e:(11)}
\end{equation}
The z-velocity is determined directly through the Doppler shifts of the
stars. The R- and T-velocities depend on the assumed Sun-Galactic centre
distance R$_{{\sun}}$. For a spatially and kinematically spherical system it
is therefore possible to derive the distance to the Galactic centre from
equation (31), without any prior assumptions on the anisotropy. The
relationship is 
\begin{equation}
(R_{{\sun}}/8{{\rm kpc}})=(\langle R\sigma _{{z}}^{{2}}\rangle _{{8}
}/\{1/3\langle R\sigma _{{R}}^{{2}}\rangle _{{8}}+2/3\langle R\sigma _{
{T}}^{{2}}\rangle _{{8}}\})^{{0.5}}\quad .  \label{e:(12)}
\end{equation}
Here $\langle $ $\rangle _{{8}}$ refers to the values calculated under the
assumption that the Galactic centre distance is 8.0 kpc, as assumed for the
proper motions in Table~1. Taking only those 32 stars for which we have all
three velocity components we find $R_{{\sun}
}=8.95\pm 1.6{{{\rm kpc}}}$. Taking all 104 proper motion stars within $
R\leq 8.8^{\prime \prime }$ and all 71 stars with z-velocities within the
same projected radius we find R$_{{\sun}}$=8.2$\pm $0.9 kpc. The specific
moment analysis in equation (32) as applied to these samples is appropriate
if the motions are completely dominated by a central point mass. In that
case $R \sigma ^{{2}}\approx ${}const and data points at different R (but
the same quality) are appropriately given the same weight. In the Galactic
centre the mass distribution is a sum of a central point mass and a
near-isothermal stellar cluster of velocity dispersion $\sigma _{{0}}$=50 to
55 km/s derived from the stellar velocities outside the sphere of influence
of the black hole (Genzel et al. 1996). It may thus be more appropriate to
subtract $\sigma _{{0}}^{{2}}$ before computing the expectation values in
equations (26) and (27). In that case we obtain $R_{{\sun}}=7.9\pm 0.85 {
{\rm kpc}}$. The difference between these two last estimates arises since
the line-of-sight velocity data are biased to a larger $\langle R\rangle $
than the proper motions so that the effect of removing $\sigma _{{0}}$ has a
larger impact on the z-velocities. This differentially decreases slightly
the distance estimate relative to that obtained for $\sigma _{{0}}=0$. All
errors do not contain a possible systematic term from deviations from
spherical symmetry.

Our analysis is in excellent agreement with other recent estimates for the
Galactic centre distance which range between 7.2 and 9.0 kpc with a best
weighted average of 8.0$\pm $0.5 kpc (see the review of Reid 1993). The
statistical uncertainty of our estimate rivals the best other methods
available for determining R$_{{\sun}}$: cluster parallaxes through H$_{{2}}
$O maser proper motions, global modeling of the Galaxy, globular cluster
dynamics, RR-Lyrae stars, Cepheids, planetary nebulae and OB stars in HII
regions (Reid 1993) and clump giant stars (Paczynski \& Stanek 1998).

\begin{figure}
\centering \includegraphics[width=1.10\columnwidth]{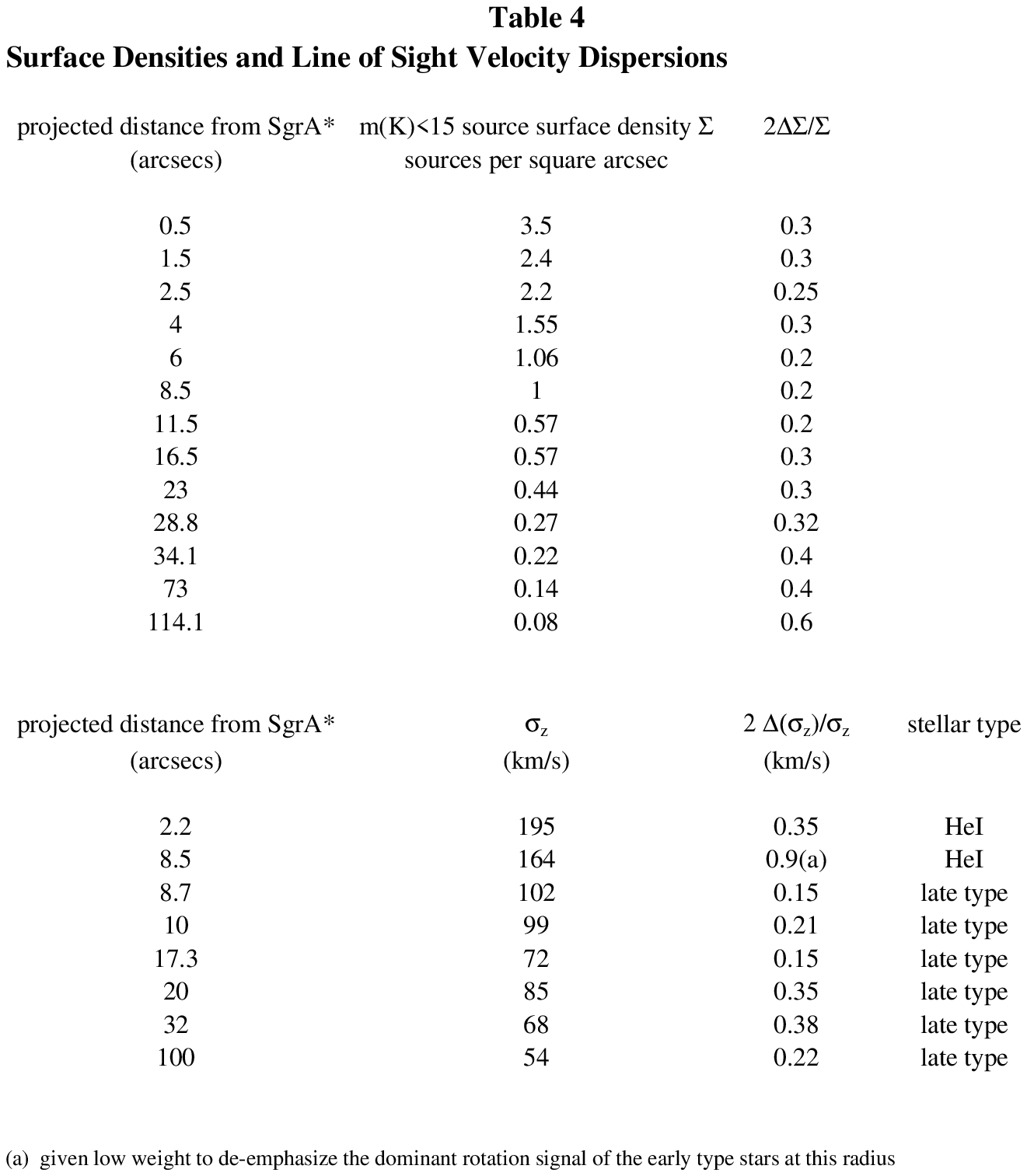}
\label{t:table1}
\end{figure}

\section{ Jeans modeling of the central mass distribution}

\begin{figure}
\centering \includegraphics[width=0.85\columnwidth]{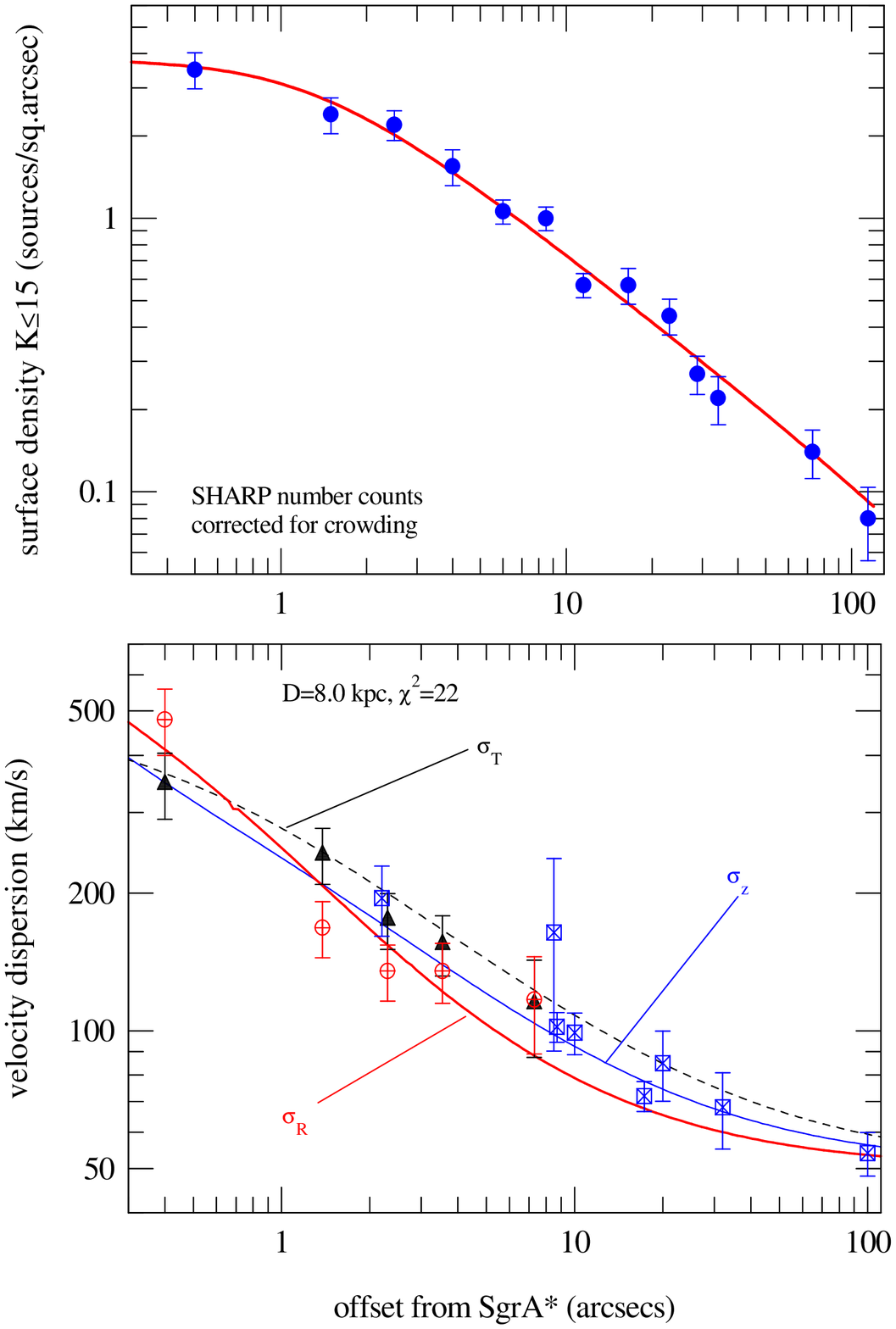}
\vskip -2cm
\caption{ Surface density and velocity dispersions as a function of
projected separation from SgrA*, along with the best fitting anisotropic
Jeans model described in section 7. The observed m(K)$\leq $15 surface
density counts and their 1$\protect\sigma $ uncertainties are shown in the
left inset (see also Table~4). These counts come from Schmitt (1995) and
have been corrected for the effects of crowding and bright stars. The
continuous curve is the model of {Eq.~(\ref{e:(13)})} with the parameters in 
Table~5, integrated along the line of sight as described in equations {~(
\ref{e:(3)})} and {~(\ref{e:(4)})}. The right inset shows observations of
the projected tangential velocity dispersionss (T: triangles, this paper, 
Table~2), projected radial velocity dispersions (R: circles, this paper, 
Table~2) and line of sight velocity dispersions (z: rectangles with
crosses, Table~4). The data for the line of sight dispersions come from
this paper (HeI stars), from Genzel et al. (1996), Haller et al. (1996),
Lindqvist et al. (1992), Sellgren et al. (1990) and McGinn et al. (1989).
The best anisotropic Jeans model integrated along the line of sight
(Table~5) gives the black, thin dashed curve for the projected tangential (T) data points,
the red, thick continuous curve for the projected radial ( R ) data points and
the blue, thin continuous curve for the line-of-sight (z) data points. The model
assumes a distance of 8 kpc and gives a total $\protect\chi ^{{2}}$ of
21.6, or a reduced $\protect\chi ^{{2}}$/N of 0.83. }
\label{f:genzel14}
\end{figure}

\begin{figure}
\centering \includegraphics[width=0.75\columnwidth]{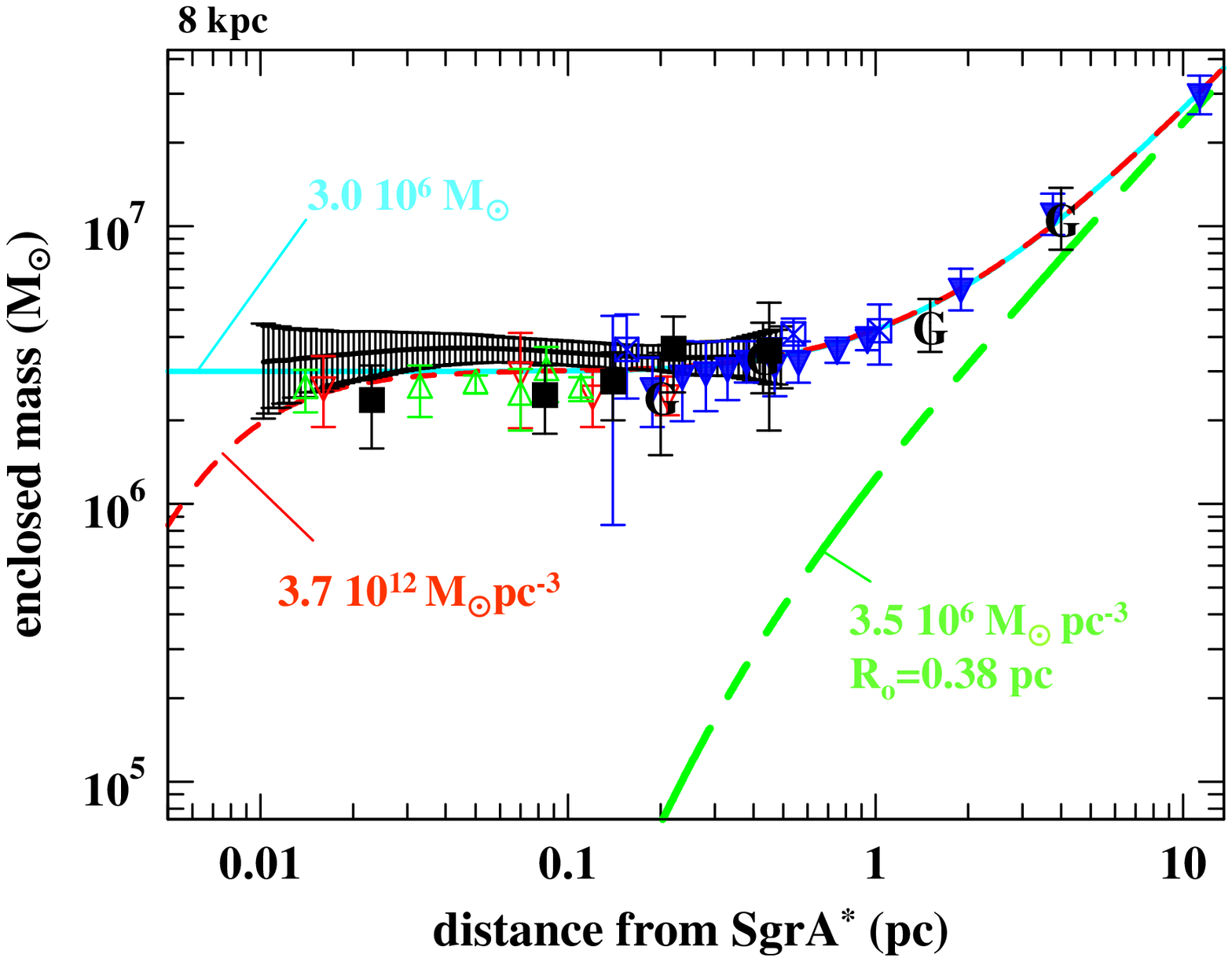}
\caption{ Mass distribution in the central 10 pc of the Galaxy obtained from
stellar and gas dynamics (for $R_{{o}}=8.0{{{\rm kpc}}}$). The different
data points (all with $\pm $1$\protect\sigma $ uncertainties) are as
follows. Bold `G' denotes mass estimates from the ionized and neutral gas
dynamics (Serabyn and Lacy 1985, G\"{u}sten et al. 1987, Lacy et al. 1991,
Herbst et al. 1993, Roberts and Goss 1993). Rectangles with crosses and
down-pointing triangles denote the isotropic mass modeling of Genzel et al.
(1996, 1997), including Jeans modeling of stellar radial velocities (early
and late type stars, filled down pointing triangles) and Bahcall-Tremaine
estimators of the NTT proper motions until 1996 (open down pointing
triangles). Open rectangles (with crosses) are Bahcall-Tremaine estimators
of the line of sight velocity data only. Open up-pointing triangles are the
Bahcall-Tremaine estimators of the 1995-1996 proper motion data of Ghez et
al. (1998). The new anisotropy independent mass estimates from the present
work are given as filled black rectangles (Leonard-Merritt estimators of the
proper motions in (Table~1, Table~2) and as large black crosses connected by a
continuous curve (Jeans model, Table~5). For comparison several model
curves are shown. The green, thick dashed curve represents the mass model for the
(visible) stellar cluster ($M/L(2\protect\mu m)=2,r_{ {\rm core}}=0.38 {
{\rm pc}}$, $\protect\rho (r=0)=3.5\times 10^{{6}}M_{{\sun}}{{\rm pc}}^{{\
-3}}$). The light blue, continuous curve is the sum of this stellar
cluster, plus a point mass of $3.0\times 10^{{6}}M_{{\sun}}$. The red,
dashed curve is the sum of the visible stellar cluster, plus an $\protect\alpha =5$
Plummer model ($\protect\rho (r)=\protect\rho (0)(1+(r/r_{{0}})^{{2}})^{{
- \protect\alpha /2}})$ of a dark cluster of central density $3.7\times
10^{{12} }M_{{\sun }}{{\rm pc}}^{{-3}}$ and $r_{{0}}=0.0058{{\rm 
pc}}$. }
\label{f:genzel15}
\end{figure}

We have also carried out a full Jeans modeling of the data set, explicitly
allowing for the anisotropy term in equation {~(\ref{e:(9)})}. It is clear
that the number of stars is still too small to unambiguously determine the
radial profiles of anisotropy and mass for all different stellar components
(and including rotation). Here we only give a simplified overall model which
is consistent with all the data. Our model proceeds from a parameterized
Ansatz for the different quantities, as described earlier in Genzel et al.
(1996), 
\begin{equation}
n(r)=2n_{{0 }}/(\pi r_{{0}}\{1+(r/r_{{0}})_{{}}^{{\alpha }
_{r}}\}),\quad \sigma _{{r}}^{{2}}(r)=\sigma _{{r0}}^{{2}}(r/r_{{0}
})_{{}}^{{-2\alpha }_{\sigma r}}+\sigma _{{0}}^{{2}}\quad {\rm and}
\quad \sigma _{{t}}^{{2}}(r)=\sigma _{{t0}}^{{2}}(r/r_{{0}})_{{}
}^{{-2\alpha }_{\sigma t}}+\sigma _{{0}}^{{2}}\quad .  \label{e:(13)}
\end{equation}

To  compare to the observed surface   density distribution $\Sigma (R)$, and
observed  velocity dispersions  $\sigma  _{{z}}(R)$, $\sigma _{{R}}(R)$  and
$\sigma _{{T}}(R)$   the  expressions   in  equation {~(\ref{e:(13)})}  were
numerically integrated along the line of sight and weighted with the density
distribution,      as   described    in    equations  {~(\ref{e:(3)})}   and
{~(\ref{e:(4)})} . The data were averaged in annuli centered around SgrA* to
yield 13 values  for  $\Sigma (R)$ between  R=0.5  and 114'', 8  values  for
$\sigma _{{z}}(R)$  between  R=2 and 100''  and  5  values each for  $\sigma
_{{R}}(R)$  and $\sigma _{{T}}(R)$ between R=0.4  and 7.3''. Best fit values
for the 8 parameters  in the expressions above  were then determined  from a
$\chi ^{{2}}$ minimization. They are listed in Table~5. The surface density
measurements come  from  number counts   with  the SHARP speckle   camera to
m(K)=15  and   are   corrected for   crowding  and  incompleteness  (Schmitt
1995). The data  points and their   statistical errors are listed in 
Table~4. The R- and T- velocity  dispersion values are from  Table~2. The line of
sight velocity dispersions  are listed in Table~4 as  well.   The two data
points at R=2.2 and 8.5'' are derived from the HeI star velocities in
Table~1. In addition we have taken late type star velocity dispersions from Genzel
et al. (1996) and references therein. The 13 surface density measurements in
Table~4  constrain the 3 parameters   of the density  distribution given in
expression  {Eq.~(\ref{e:(13)})}   very  well.  Likewise    the  18 velocity
dispersion measurements  also give good constraints  on the 5  parameters of
the dispersion expressions in {Eq.~(\ref{e:(13)})}.

{Fig.~(\ref{f:genzel14})} shows the surface density and velocity data, along
with  the best fit model  whose parameters are given  in Table~5. Table~5
also lists   the  best anisotropic mass  model,   along with the logarithmic
gradients and the $\beta $ -anisotropy parameter that were used in the Jeans
equation ({Eq.~(\ref{e:(9)} )}) to derive that mass model.

\begin{figure}
\centering \includegraphics[width=1.00\columnwidth]{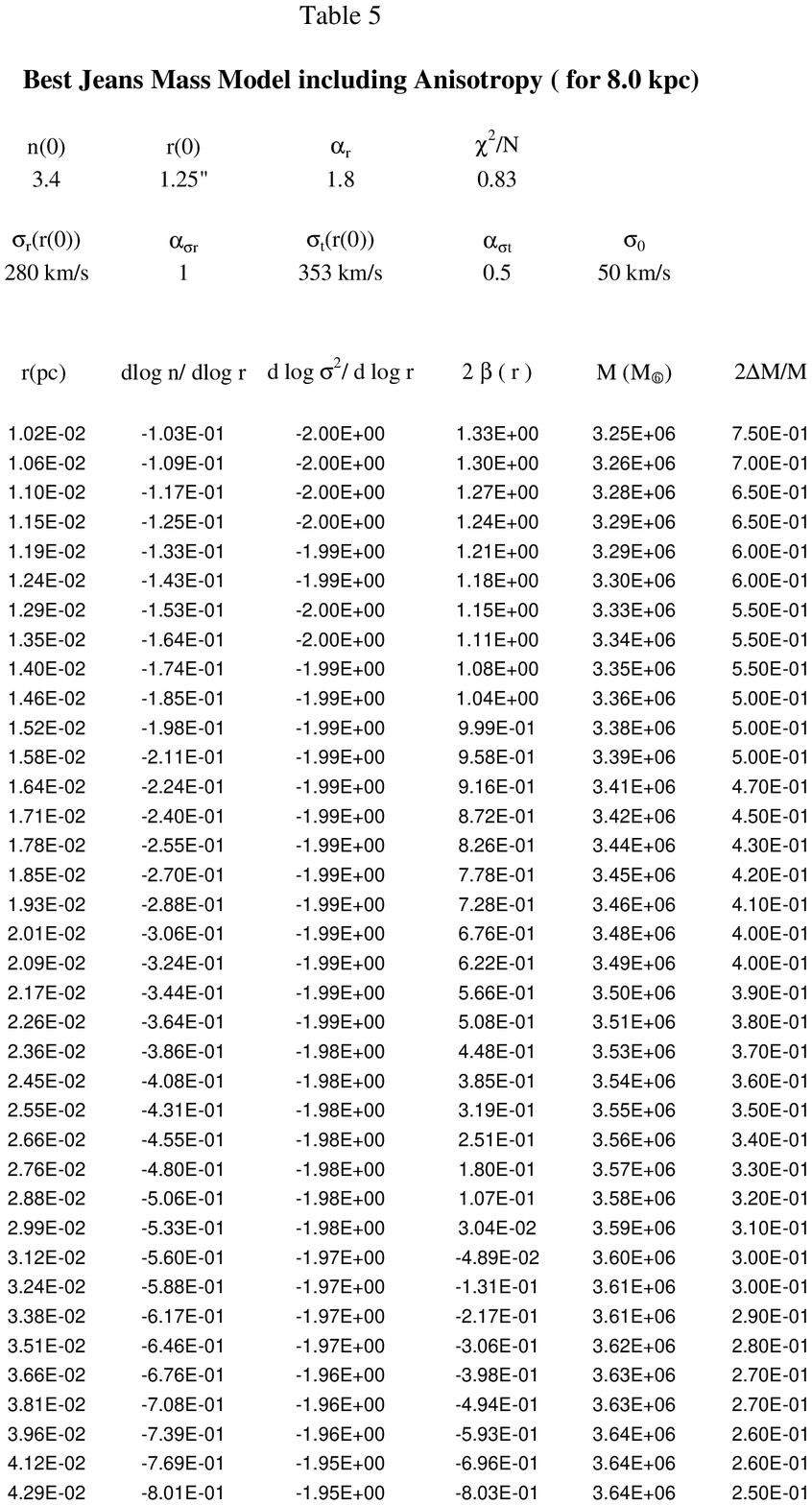}
\label{t:table1}
\end{figure}
\begin{figure}
\centering \includegraphics[width=1.00\columnwidth]{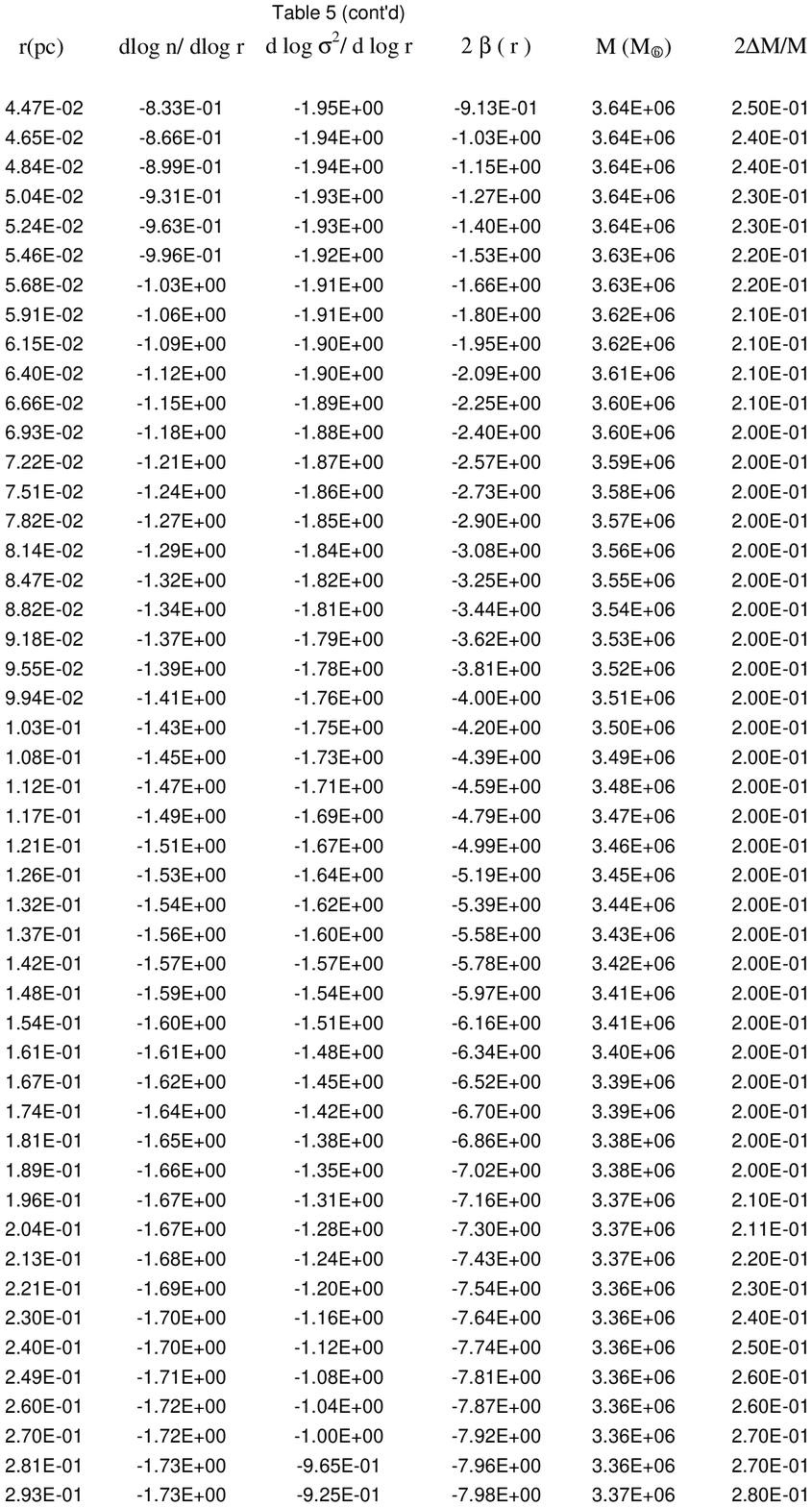}
\label{t:table1}
\end{figure}
\begin{figure}
\centering \includegraphics[width=1.00\columnwidth]{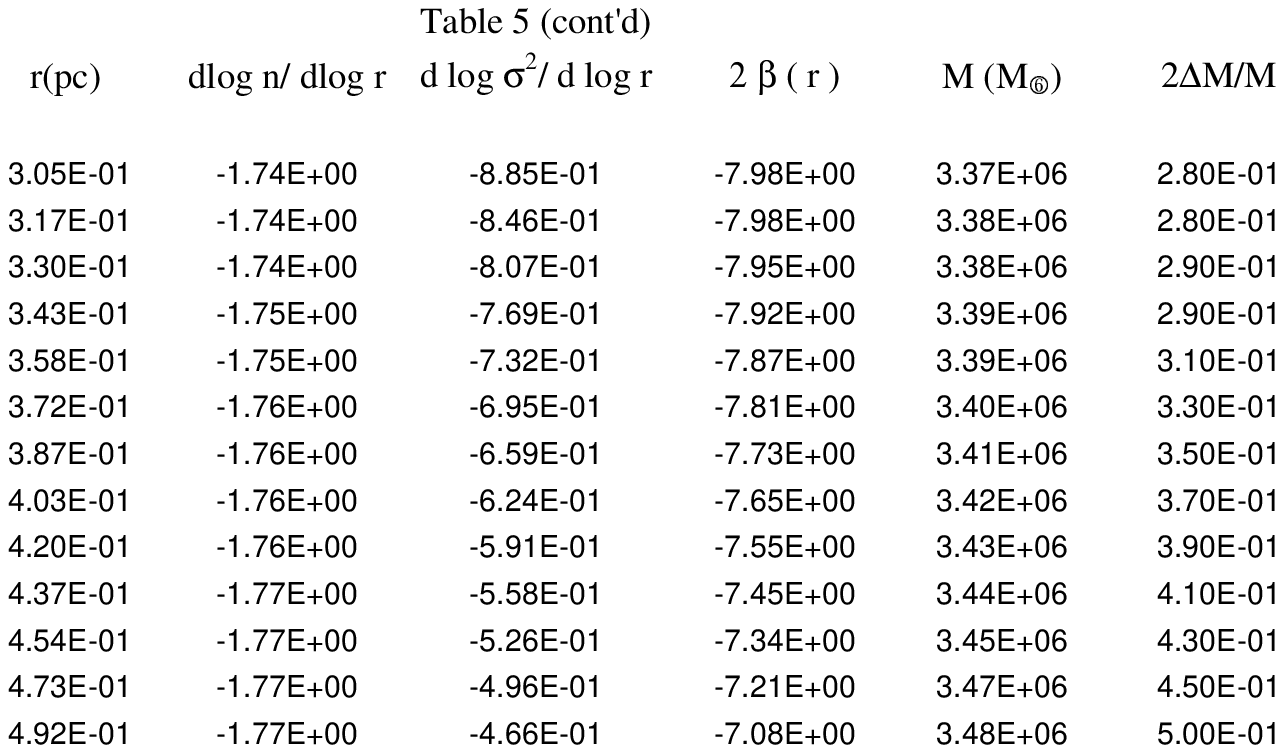}
\label{t:table1}
\end{figure}

\section{Discussion and Conclusions}

The connected black crosses in {Fig.~(\ref  {f:genzel15})}  depict the 
 mass distribution  obtained from the Jeans-model with anisotropy (and its
(1$\sigma $) uncertainty). For comparison we also show the LM-mass estimators
from Table~2, BT estimators for the  z- and proper motions  (this paper,
Genzel et al.  1996, 1997, Ghez et al. 1998), the isotropic Jeans mass model
of Genzel et al.(1996) and  a few of the mass  estimates determined from the
gas motions (Guesten et al.  1987, Serabyn and Lacy 1985,  Lacy et al. 1991,
Herbst et al.  1993, Roberts et al. 1993). The  mass inside of the innermost
bin ($r=0.01{ {\rm pc}}$)  of our best   Jeans model is $3.25\times  10^{{6}
}M_{\sun }$. It is consistent with the Leonard-Merritt mass estimator of the
entire   proper motion sample   inside $5^{\prime \prime },M_{{LM} }=2.9\pm
0.4\times 10^{{  6}}M_{\sun }$ (Table~2,  see also Section 6). When 
corrected for the bias
discussed above that mass becomes about 
$2.6-2.8 \times 10^{{  6}}M_{\sun }$.  Systematic effects and the
method of modelling dominate  the accuracy to which  the
central mass can be  determined. In \Fig{genzel15}  we plot a central mass of
$3.0 \times  10^{6} M_{\sun}$. This value is a compromise
between the bias-corrected LM-estimate and the Jeans estimate.
Its  overall (systematic plus statistical)
uncertainty is $\pm 0.5\times 10^{6} M_{\sun}$. 
It is reassuring that the results of our
simple   anisotropic  modeling  and  of  previous  isotropic  models are  in
good agreement within the   respective  errors. {\bf {\it   Our results
confirm that the mass distribution is  flat between $0.01$ and $0.5{{\rm pc
}}\/$}}.

Nonetheless, there are still significant uncertainties in this analysis. 

(i)
The parametric form of the model  fitted to the data is  not unique. 

(ii) We
have so far not distinguished between early  and late type  stars. Yet it is
fairly clear  that   early  and late   type stars   have  different  spatial
distributions and kinematics  (see Plate 1  in Genzel et  al. 1996).  In the
Jeans analysis, we  require the  density distribution  and kinematics  of an
equilibrium  tracer   population. The proper     motions  stars are  heavily
influenced by spatial  selection biases; thus it  is not  appropriate to use
their inferred number density distribution in the Jeans equation. Their role
is to  provide local  velocity   measurements for the  population that  they
represent.  The  early-type    stars  contribute  much to  these   kinematic
measurements; if  they  are more  centrally  concentrated  than the  overall
population measured by the SHARP number counts, this will have the effect of
underestimating the central mass. 

(iii) The version of the Jeans equation we
have used in  expression  {~(\ref{e:(9)})} neglects  rotation. Yet  we  have
discussed above  the strong  evidence for coherent   motion of the  HeI star
cluster. The late  type stars have  only a small overall rotation. Including
rotation and  distinguishing in  the  analysis between  late and early  type
stars would thus  be desirable (as  in Genzel et  al. 1996 for the isotropic
case). Unfortunately  this is not possible, because  of  the large number of
free  parameters (3 more for  density distribution, approximately 8 more for
velocity distribution) and  the relatively poor constraints   on a number  of the
parameters. There are no  early type stars outside 11'',  there are very few
late type stars inside   5''and the accuracy  of  the proper motion  R-  and
T-velocity dispersions is  low if all proper motions  without a stellar type
identification are discarded. We have run models  with explicit inclusion of
rotation  but found it  to be overall a  poorer  fit than the models without
rotation. To deliberately deemphasize the rotation signature of the HeI star
cluster we  have arbitrarily given   the z-velocity dispersion value  at  $
R=8.5^{\prime\prime}$ (Table~4) a low weight.  {
\vskip0.21cm}

The central   dark mass  concentration  is  most likely a  point   mass. Any
configuration other than a  point mass must have a  central density of $\rho
(0)\geq 3.7 \times 10^{{12} }M_{\sun }{{\rm pc}}^{{-3}}$and a core radius of
$r_{{0}}\leq 5.8$ milli-parsec . For this estimate we have adopted
 a Plummer model with
a  density profile that  decreases as
$r^{{-5}}$ outside  of the core   radius. In a configuration 
with a point  mass  and the visible  stellar
cluster   $(\rho   _{{\ast   }}(0)=3.5\times   10^{{6}}M_{\sun }{{\rm   pc}
}^{{-3}},\alpha =1.8,r_{{0} }=0.17{{\rm pc}}$) as the two main components of
the mass distribution any additional mass within $\approx 0.2-0.5{{\rm pc}}$
of SgrA* must be less than $\approx 1\times 10^{{6} }M_{\sun }$, or 32\% of
the point mass.  If one takes the LM-mass  distribution  instead (Table~2),
that limit  would be between $1.1$ and  $ 2.2  \times 10^{{6}}M_{\sun }.$
Backer (1996) has shown that the  proper motion of  SgrA* itself is $\leq 16
{\rm km\,s}^{-1}$, or  fifty  to one  hundred  times smaller  than  the fast
moving stars in its  vicinity. Thus the mass enclosed  within the radio size
of  SgrA* ($r\leq  1AU$) is $\geq  10^{{3} }$  or $\geq 10^{{5}}M_{\sun }$,
depending on whether  the radio source  is in momentum or energy equilibrium
with the fast  moving stars (Genzel et al.   1997, Reid 1999). Even the more
conservative  of these two  limits  implies a central   density in excess of
$10^{{18}}M_{\sun }{{\rm pc}}^{{-3}}$.{
\vskip0.21cm}

Our  results confirm   and  strengthen  recent  work   on  the central  mass
distribution  (cf. Eckart and Genzel 1996,  1997, Genzel et al. 1997, Genzel
and Eckart 1997 , Ghez et al. 1998). From these papers  and from Maoz (1998)
it appears   that  the  most likely   configuration   of  the central   mass
concentration is  a massive,  but currently  inactive  black hole. With  the
parameters given above any dark  cluster of stellar remnants (neutron stars,
stellar  black   holes), low  luminosity   stars  (e.g.  white   dwarfs)  or
sub-stellar  objects would  have  a lifetime less  than $\approx {}10^{{7}}$
years.  This is much   smaller than the ages  of  most of  the stars in  the
Galactic centre, requiring that we happen  to observe the Galactic centre in
a highly  improbable, special period.   In addition,  the very  steep  outer
density distribution of such a dark cluster implied by the mass distribution
in {Fig.~(\ref{f:genzel15})} $(\rho \approx {}r^{{-\alpha }}$with $\alpha
\geq 5)$ is inconsistent with any known observed dynamical system. It is
also inconsistent with the results of physical models, including those of
core-collapsed clusters (see the discussion of Genzel et al. 1997). Maoz
(1998) points out that the only possible - albeit highly implausible -
alternatives to a central black hole are a concentration of heavy bosons and
a compact cluster of light $(\leq 0.005M_{\sun })$ `mini'-black holes.

\vskip 1cm {\bf {\it Acknowledgements. \/}}{\sl We would like to thank David
Merritt for making us aware  of the mass  estimators  first laid out in  his
paper with Peter Leonard in 1989, as well as for comments on the
present manuscript. We  also appreciate the willingness of the
ESO Director General and his staff  to let us  bring the SHARP camera to the
NTT in 1997 and 1998. We thank Niranjan Thatte, Alfred Krabbe, Harald Kroker
and the entire  MPE-3D team for their help  with the 3D observations  on the
ESO-MPG 2.2m  and for developing the  new data reduction tools  necessary to
analyse the data set  presented here. We  are also grateful to the  NTT-team
and  especially to U.Weidenmann  and   H.Gemperlein for  their interest  and
technical support of SHARP at the NTT.  CP would like to thank E.~Thi\'ebaut
for many  valuable  discussions. We thank Lowell Tacconi-Garman and
Linda Tacconi for comments on and help with the
manuscript.  We are grateful to the
referee for his valuable and helpful comments.
Funding from  the  Swiss NF  is gratefully
acknowledged. \/}


\end{document}